\documentclass[preprint,5p,times, twocolumn]{elsarticle}

\usepackage{amssymb}
\usepackage{amsmath}
\usepackage[margin=0pt,font=small,labelfont=bf,labelsep=colon]{caption}
\usepackage{algorithm}
\usepackage{algpseudocode}
\usepackage{listings}
\usepackage{color}
\usepackage{textcomp}
\usepackage{url}
\definecolor{listinggray}{gray}{0.9}
\definecolor{lbcolor}{rgb}{0.9,0.9,0.9}
 \biboptions{sort&compress}

\journal{Computer Physics Communication}

\begin{document}

\begin{frontmatter}

\author{Zheyong Fan\corref{cor1}}
\ead{zheyong.fan@aalto.fi}
\cortext[cor1]{Corresponding author}
\author{Andreas Uppstu}
\author{Topi Siro}
\author{Ari Harju}

\title{Efficient linear-scaling quantum transport calculations
       on graphics processing units and applications on electron
       transport in graphene}

\address{COMP Centre of Excellence and Helsinki Institute of Physics,
Department of Applied Physics, Aalto University, Helsinki, Finland}

\begin{abstract}
We implement, optimize,
and validate the linear-scaling Kubo-Greenwood quantum transport
simulation on graphics processing units by examining resonant 
scattering in graphene.  We consider two practical
representations of the Kubo-Greenwood formula: a Green-Kubo
formula based on the velocity auto-correlation and an
Einstein formula based on the mean square displacement.
The code is fully implemented on graphics processing units
with a speedup factor of up to 16 (using double-precision)
relative to our CPU implementation.
We compare the kernel polynomial method and the Fourier transform method
for the approximation of the Dirac delta function and conclude that
the former is more efficient.
In the ballistic regime, the Einstein formula can produce
the correct quantized conductance of one-dimensional graphene nanoribbons
except for an overshoot near the band edges.
In the diffusive regime, the Green-Kubo and the Einstein formalisms
are demonstrated to be equivalent.
A comparison of the length-dependence of the conductance in the
localization regime obtained by the Einstein formula with
that obtained by the non-equilibrium Green's function method reveals the
challenges in defining the length in the Kubo-Greenwood formalism
at the strongly localized regime.
\end{abstract}


\end{frontmatter}

\section{Introduction}
\label{Introduction}

Quantum simulations are very important tools to study
transport phenomena in the nanoscale, both for electrons and phonons. There
are mainly two numerical approaches for quantum transport simulations,
one is the widely used non-equilibrium Green's
function (NEGF) method \cite{datta1995} and the other is
the Kubo-Greenwood method \cite{kubo1957, greenwood1958}. Both methods have been widely
used to study the electronic properties of graphene, a two-dimensional sheet of carbon atoms
 \cite{geim2007, neto2009}. Despite this, the field of electronic transport in graphene has
 remained very actively debated.

So far, the NEGF method has been mostly used to simulate relatively
small systems, due to the cubic scaling of the computational effort
associated with matrix inversion. Although an efficient iterative method
\cite{sancho1985}
enables the simulation of very long systems, this method is still
restricted to studying quasi-one-dimensional (1D) systems, such as
carbon nanotubes and graphene nanoribbons (GNRs). The application
of the NEGF method to realistically sized two-dimensional
(2D) graphene is still not feasible.

In contrast, for the Kubo-Greenwood method, a real-space
linear-scaling method
has been developed \cite{mayou1988,mayou1995,roche1997,triozon2002}
and used to study transport properties of both
quasi-1D systems \cite{triozon2004,markussen2006,ishii2010}
and 2D graphene sheets
\cite{lherbier2008a,lherbier2008b,laissardiere2011,
leconte2011,radchenko2012,lherbier2012,tuan2013,cresti2013}.
Moreover, this method has been generalized to studying thermal
conductivity \cite{li2011}.
Besides the real-space Kubo
method \cite{mayou1988,mayou1995,roche1997,triozon2002},
which expresses the conductivity as a time-derivative
of the mean square displacement,
another seemingly different approach \cite{yuan2010},
which expresses the conductivity as a time-integration
of the velocity auto-correlation function, has also been used
to study the electronic transport properties of large-scale
single-layer \cite{yuan2010} and multi-layer
\cite{yuan2010b} graphene sheets, and disordered graphene
antidot lattices \cite{yuan2013}.

Although both of the above methods are based
on the Kubo-Greenwood formula, no connection has been made
between them. One of our purposes is to identify the time-derivative
approach and the time-integration approach as
an Einstein relation and the corresponding Green-Kubo relation,
and demonstrate their equivalence numerically.
Furthermore, a thorough validation of
Kubo-Greenwood formula based quantum transport methods
for all the transport regimes is also absent.
We thus aim to perform a comprehensive evaluation
of the applicability of the linear-scaling Kubo-Greenwood
quantum transport simulation method for all three transport
regimes: the ballistic, diffusive, and localized regimes.

To achieve the above, we find that
an efficient implementation
is very desirable. Despite the linear-scaling
nature of these numerical methods,
they are still computationally demanding in most cases.
Nowadays, the use of graphics processing units (GPUs)
have played a more and more important role in
computational physics; finding the solutions to many
problems in computational physics has become impressively
accelerated by using a single or
multiple GPUs \cite{ari2012}.
In this work, we consider the implementation of
the Kubo-Greenwood quantum transport simulation on
the GPU, with a unified treatment of the various involved
theoretical formalisms and numerical techniques.
We will evaluate the performance and correctness of our implementation,
as well as the applicability of the method itself.

This paper is organized as follows. In section \ref{section:formalism},
we present the theoretical background of the Kubo-Greenwood formula
and the Green-Kubo and Einstein relations which are both
derived. In section \ref{section:implementation},
we give a detailed discussion of the involved
numerical techniques and their GPU implementations.
After making a performance evaluation
in section \ref{section:Performance}, we thoroughly
evaluate the computational method in different transport
regimes in section \ref{section:validation}.
Section \ref{section:Conclusion} concludes.

\section{Theoretical formalism}
\label{section:formalism}

The Kubo-Greenwood formula \cite{greenwood1958} for DC conductivity
$\sigma^{\textmd{KG}}_{\mu \nu}(E)$ as a function of the energy $E$
at zero temperature is
\begin{equation}
\sigma^{\textmd{KG}}_{\mu \nu}(E) = \frac{2 \pi \hbar e^2}{\Omega}
            \textmd{Tr}\left[V_{\mu} \delta(E-H) V_{\nu} \delta(E-H)\right].
\end{equation}
where $\hbar$ is the reduced Plank constant, $e$ is the electron charge,
$\Omega$ is the system volume, $V_{\mu}$ is the velocity operator
in the $\mu$-direction, $H$ is the Hamiltonian of the system, and
Tr denotes the trace. The factor of two results from spin degeneracy.
For simplicity, we only consider transport along one direction.
Then, the above formula can be simplified to be
\begin{equation}
\sigma^{\textmd{KG}}(E) = \frac{2 \pi \hbar e^2}{\Omega}
            \textmd{Tr}\left[V \delta(E-H) V \delta(E-H)\right].
\end{equation}

By Fourier transforming one of the $\delta$ functions in the above formula,
\begin{equation}
 \delta(E - H) = \frac{1}{2 \pi \hbar} \int_{-\infty}^{+\infty} dt
                 e^{i (E - H) t / \hbar},
\end{equation}
we have
\begin{equation}
\sigma(E) = \frac{e^2}{\Omega} \int_{-\infty}^{+\infty} dt
            \textmd{Tr}\left[e^{i E t / \hbar} V e^{-i H t / \hbar} V
            \delta(E-H) \right],
\end{equation}
or equivalently,
\begin{equation}
\sigma(E) = \frac{e^2}{\Omega} \int_{-\infty}^{+\infty} dt
            \textmd{Tr}\left[e^{i H t / \hbar} V e^{-i H t / \hbar} V
            \delta(E-H) \right].
\end{equation}
due to the remaining $\delta$ function.
Through a change of variables, $t \rightarrow -t$,
we get the following Green-Kubo formula \cite{green1954,kubo1957},
which expresses the
running electrical conductivity (REC)
as a time-integration of the velocity auto-correlation (VAC) $C_{vv}(E, t)$,
\begin{equation}
\label{equation:REC_GK}
\sigma^{\textmd{GK}}(E, t) = e^2 \rho(E) \int_0^{t} C_{vv}(E,t) d t,
\end{equation}
\begin{equation}
C_{vv}(E, t) = \frac{\textmd{Tr}\left[ \frac{2 }{\Omega}\delta(E-H)
                \left(V(t)V + VV(t)\right)/2\right]}
              {\textmd{Tr}\left[\frac{2 }{\Omega}\delta(E-H)\right]},
\end{equation}
\begin{equation}
\rho(E) = \textmd{Tr}\left[\frac{2 }{\Omega}\delta(E-H)\right],
\end{equation}
where $V(t) = U^{\dagger}(t) V U(t) = e^{iHt/\hbar}Ve^{-iHt/\hbar}$
is the velocity operator in the Heisenberg representation,
and $\rho(E)$ the density of states (DOS).
The Green-Kubo relation constitutes essentially the formalism used
by Yuan \textit{et al.} \cite{yuan2010,yuan2010b}.

For a specific Green-Kubo formula, there is generally
a corresponding Einstein formula.
By integrating the Green-Kubo formula, we obtain the following
Einstein formula, which expresses the REC as a  time-derivative
of the mean square displacement (MSD) $\Delta X^2(E, t)$,
\begin{equation}
\label{equation:REC_E1}
\sigma^{\textmd{E1}}(E, t) = e^2 \rho(E) \frac{d}{2 d t}
            \Delta X^2(E, t),
\end{equation}
\begin{equation}
\Delta X^2(E, t) =
\frac{\textmd{Tr}\left[ \frac{2}{\Omega}\delta(E-H)
\left(X(t) - X\right)^2\right]}
{\textmd{Tr}\left[\frac{2 }{\Omega}\delta(E-H)\right]},
\end{equation}
where $X(t) = U^{\dagger}(t) X U(t)$ is the position operator in the
Heisenberg representation.
An alternative definition, in which
the derivative in the above equation is replaced by a division,
\begin{equation}
\label{equation:REC_E2}
 \sigma^{\textmd{E2}}(E, t) = e^2 \rho(E) \frac{\Delta X^2(E, t)}{2t},
\end{equation}
is frequently used, since it gives smoother curves for the REC
than $\sigma^{\textmd{E1}}(E, t)$ does.
The above Einstein relation is exactly the
real-space Kubo method \cite{mayou1988,mayou1995,roche1997,triozon2002}.

We will demonstrate the equivalence of the Green-Kubo formalism
and the Einstein formalism numerically. Specifically, we will show
that $\sigma^{\textmd{E1}}(E, t)$ and $\sigma^{\textmd{GK}}(E, t)$
are equivalent, while $\sigma^{\textmd{E2}}(E, t)$ deviates
from the other two to some degree.

By going from the Kubo-Greenwood formalism to the Green-Kubo or
the Einstein formalism,
the conductivity becomes a function of not only the energy $E$, but
also the correlation time $t$. Usually,
one takes the following large time limit:
\begin{equation}
\label{equation:limit}
 \sigma^{\textmd{KG}}(E) = \lim_{t \to \infty} \sigma^{\textmd{GK}}(E, t)
                         = \lim_{t \to \infty} \sigma^{\textmd{E1}}(E, t).
\end{equation}
However, the convergence of this limit is only ensured for diffusive
transport, in which case the VAC decays to zero and
the MSD becomes proportional to $t$, resulting in a converged REC.
For ballistic transport, the VAC oscillates around a fixed
value and the MSD increases quadratically with increasing $t$,
resulting in a divergent REC. In the localized regime,
the VAC develops negative values and the slope of the MSD
decreases, resulting in a decaying REC.

In this paper, we take graphene
as our test system. We use $N_x$ to represent the number of
dimer lines located along the zigzag edge and $N_y$
to represent the number of zigzag-shaped chains across
the armchair edge. Thus, an
$N_x \times N_y$ graphene sample has $N=N_x N_y$ carbon atoms,
and the lengths in the zigzag and armchair directions are
$L_x = \sqrt{3} N_x  a / 2$ and $L_y = 3 N_y a / 2$, respectively,
where $a=0.142$ nm is the carbon-carbon bond length used.
For 2D graphene, periodic boundary conditions
are applied in both directions; for quasi-1D
armchair graphene nanoribbon (AGNR) and zigzag graphene
nanoribbon (ZGNR), we use periodic boundary conditions along the transport
(longitudinal) direction, and non-periodic boundary conditions
along the perpendicular direction.

We use a nearest-neighbor $p_z$ orbit tight-binding Hamiltonian
for pristine systems:
\begin{equation}
 H = \sum_{\langle mn \rangle} H_{mn} |m\rangle \langle n|
   = -\sum_{\langle mn \rangle}\gamma_{0} |m\rangle \langle n|,
\end{equation}
where the hopping parameter $\gamma_{0}$ is chosen to be 2.7 eV.
With this notation,
the position and velocity operators can be expressed as
\begin{equation}
 X = \sum_{m} X_m |m\rangle \langle m|,
\end{equation}
\begin{equation}
 V = \frac{i}{\hbar} [H,X]
= \frac{i}{\hbar}\sum_{\langle mn \rangle}
 (X_n-X_m) H_{mn} |m\rangle \langle n|.
\end{equation}

We also consider systems with random single vacancies, which are modeled
by removing carbon atoms randomly according to the prescribed defect
concentrations. The defect concentration $n$ is determined by the
system size $N$ and the number of vacancies $N_v$ as $n=N_v/N$.

\section{Numerical implementation}
\label{section:implementation}

\subsection{Numerical approximations}

Based on the discussion of the last section, we see that
the quantities that need to be calculated are $\rho (E)$,
$\rho (E) C_{vv}(E, t)$, and $\rho (E) \Delta X^2(E, t)$.
To facilitate the numerical calculation,
we firstly rewrite $\rho (E) C_{vv}(E, t)$ and $\rho (E) \Delta X^2(E, t)$
in the following symmetric forms (using the cyclic properties of the trace):
\begin{equation}
 \rho (E) C_{vv}(E, t) =
 \frac{2}{\Omega} \textmd{Re}
 \left[ \textmd{Tr} \left[U(t) V\delta(E - H) U(t)^{\dagger}V\right] \right]
\end{equation}
\begin{equation}
 \rho (E) \Delta X^2(E, t) =
 \frac{2}{\Omega} \textmd{Tr}
 \left[[X, U(t)]^{\dagger}\delta(E - H) [X, U(t)]\right].
\end{equation}
The reason for this will be apparent
when we consider the GPU-implementation.
To achieve linear-scaling, we have to make three approximations
presented below.

\subsubsection{Approximation of the trace}

The first approximation is to use a random vector $|\phi\rangle$
to evaluate the trace \cite{weibe2006}:
\begin{equation}
 \textmd{Tr}\left[ A \right] \approx \langle \phi | A |\phi \rangle,
\end{equation}
where $A$ is an arbitrary matrix operator, and $|\phi\rangle$
is normalized to the matrix dimension $N$,
$\langle \phi | \phi \rangle = N$.
With this approximation, we have
\begin{equation}
\label{equation:trace_DOS}
 \rho(E) \approx \frac{2}{\Omega} \langle \phi | \delta(E - H) |\phi \rangle,
\end{equation}
\begin{equation}
\label{equation:trace_VAC}
 \rho (E) C_{vv}(E, t) \approx
 \frac{2}{\Omega} \textmd{Re}
 \left[ \langle\phi|U(t)V \delta(E - H) U(t)^{\dagger}V|\phi \rangle \right],
\end{equation}
\begin{equation}
\label{equation:trace_MSD}
 \rho (E) \Delta X^2(E, t) \approx
 \frac{2}{\Omega}
 \langle\phi|[X, U(t)]^{\dagger} \delta(E - H)[X, U(t)]|\phi \rangle.
\end{equation}
The error introduced by this approximation decreases with increasing $N$.
For a given $N$, the accuracy can also be increased by using
a higher number of random vectors. Quantitatively, the relative error is of
order $O(1/\sqrt{N_r N})$ \cite{weibe2006}, where $N_r$ is the
number of random vectors.

\subsubsection{Approximation of the $\delta$ function}

The second approximation is related to the $\delta$ function. There are
various kinds of methods to approximate this, including
the Lanczos recursion method (LRM) \cite{haydock1972,haydock1975}, the
Fourier transform method (FTM) \cite{feit1982,hams2000},
and the kernel polynomial method (KPM)
\cite{weibe2006}. The LRM and the KPM has been compared in
Ref. \cite{weibe2006}. In this work, we use the FTM and the KPM
and give a comparison of them.

In the FTM \cite{feit1982,hams2000},
the $\delta$ function is approximated by a truncated
discrete Fourier series expansion, and we can rewrite
Eqs. (\ref{equation:trace_DOS}-\ref{equation:trace_MSD})
as
\begin{equation}
\label{equation:FTM_DOS}
 \rho(E) \approx \frac{\Delta \tau}{\pi \hbar \Omega}
 \sum_{n=-N_m}^{+N_m} w_n e^{iEn\Delta \tau/\hbar} F_n^{\textmd{DOS}},
\end{equation}
\begin{equation}
\label{equation:FTM_VAC}
 \rho (E) C_{vv}(E, t) \approx
 \frac{\Delta \tau}{\pi \hbar \Omega}
 \sum_{n=-N_m}^{+N_m} w_n \textmd{Re}
 \left[ e^{iEn\Delta \tau/\hbar} F_n^{\textmd{VAC}} (t) \right],
\end{equation}
\begin{equation}
\label{equation:FTM_MSD}
 \rho (E) \Delta X^2(E, t) \approx
 \frac{\Delta \tau}{\pi \hbar \Omega}
 \sum_{n=-N_m}^{+N_m} w_n e^{iEn\Delta \tau/\hbar} F_n^{\textmd{MSD}} (t),
\end{equation}
where $F_n^{\textmd{DOS}}$, $F_n^{\textmd{VAC}}(t)$,
and $F_n^{\textmd{MSD}}(t)$ are the Fourier moments:
\begin{equation}
\label{equation:F_n_DOS}
 F_n^{\textmd{DOS}} \approx \langle \phi | U(n \Delta \tau) |\phi \rangle,
\end{equation}
\begin{equation}
\label{equation:F_n_VAC}
 F_n^{\textmd{VAC}}(t)
 \approx \langle\phi|U(t)V U(n \Delta \tau) U(t)^{\dagger}V|\phi \rangle,
\end{equation}
\begin{equation}
\label{equation:F_n_MSD}
 F_n^{\textmd{MSD}}(t) \approx
 \langle\phi|[X, U(t)]^{\dagger} U(n \Delta \tau) [X, U(t)]|\phi \rangle.
\end{equation}
Note that a window function should be applied before
performing the Fourier transform to suppress the unwanted
Gibbs oscillation. Usually, a Hanning window
\begin{equation}
 w_n = \frac{1}{2} \left[1 + \cos\left(\frac{\pi n}{N_m+1}\right)\right]
\end{equation}
is used \cite{feit1982}.
We will discuss the choice of the time step $\Delta \tau$ used in
the above Fourier transforms when we compare the relative performance
of the FTM and the KPM in the next section.

In the KPM \cite{weibe2006},
the $\delta$ function is approximated by a truncated
Chebyshev polynomial expansion, and we can rewrite
Eqs. (\ref{equation:trace_DOS}-\ref{equation:trace_MSD})
as
\begin{equation}
\label{equation:KPM_DOS}
 \rho(E) \approx \frac{2}{\pi \Omega \sqrt{1-E^2}}
 \sum_{n=0}^{N_m-1} g_n (2-\delta_{n0}) T_n(E) C_n^{\textmd{DOS}},
\end{equation}
\begin{equation}
\label{equation:KPM_VAC}
 \rho (E) C_{vv}(E, t) \approx
 \frac{2}{\pi \Omega \sqrt{1-E^2}}
 \sum_{n=0}^{N_m-1} g_n (2-\delta_{n0}) \textmd{Re}
 \left[T_n(E) C_n^{\textmd{VAC}} (t) \right],
\end{equation}
\begin{equation}
\label{equation:KPM_MSD}
 \rho (E) \Delta X^2(E, t) \approx
 \frac{2}{\pi \Omega \sqrt{1-E^2}}
 \sum_{n=0}^{N_m-1} g_n (2-\delta_{n0}) T_n(E) C_n^{\textmd{MSD}} (t),
\end{equation}
where $T_n(E)$ is the $n$th order Chebyshev polynomial of the first kind
and $C_n^{\textmd{DOS}}$, $C_n^{\textmd{VAC}}(t)$,
and $C_n^{\textmd{MSD}}(t)$ are the Chebyshev moments:
\begin{equation}
\label{equation:C_n_DOS}
 C_n^{\textmd{DOS}} \approx \langle \phi | T_n(H) |\phi \rangle,
\end{equation}
\begin{equation}
\label{equation:C_n_VAC}
 C_n^{\textmd{VAC}}(t) \approx
 \langle\phi|U(t)V T_n(H) U(t)^{\dagger}V|\phi \rangle,
\end{equation}
\begin{equation}
\label{equation:C_n_MSD}
 C_n^{\textmd{MSD}}(t) \approx
 \langle\phi|[X, U(t)]^{\dagger} T_n(H) [X, U(t)]|\phi \rangle.
\end{equation}
Similarly, a damping factor should be applied before performing the
Chebyshev summation in order to suppress the Gibbs oscillation.
Usually, the Jackson damping \cite{weibe2006}
\begin{equation}
 g_n = \left(1 - n \alpha \right) \cos\left(\pi n \alpha \right)
     + \alpha \sin\left(\pi n \alpha \right) \cot\left(\pi \alpha\right),
\end{equation}
where $\alpha = 1/(N_m+1)$ is used.
Note that the above Chebyshev expansions assume that $H$
has been scaled and shifted \cite{weibe2006}
so that the spectrum lies in the interval
$[-1, 1]$ .

Both the Fourier and the Chebyshev moments can be evaluated iteratively.
Detailed algorithms will be presented when we consider the GPU-implementation.

\subsubsection{Approximation of the time-evolution}

The third approximation is to evaluate the application of the
time-evolution operators on state vectors using a
finite-term polynomial expansion.
From the discussion above, we see that
there are three kinds of time-evolution
operators: $U(\Delta t)$, $U(\Delta t)^{\dagger} = U(-\Delta t)$,
and $[X, U(t)]$.
Their operations can be evaluated very accurately and efficiently
in a linear-scaling way by using the
Chebyshev polynomial expansion \cite{ezer1984,fehske2009}:
\begin{equation}
\label{equation:cheb_1}
 U(\pm \Delta t)
 \approx \sum_{m=0}^{N_p-1} (2-\delta_{0m}) (\mp i)^m
         J_m\left( \frac{\Delta t}{\hbar} \right)
         T_m\left( H \right) ,
\end{equation}
\begin{equation}
\label{equation:cheb_2}
 [X, U(\Delta t)]
 \approx \sum_{m=0}^{N_p-1} (2-\delta_{0m}) (-i)^m
J_m\left( \frac{\Delta t}{\hbar} \right)  [X, T_m(H)],
\end{equation}
where $J_m\left( \frac{\Delta t}{\hbar} \right)$
is the $m$th order Bessel function of the first kind.
Time-evolution of quantum states has also been considered
with regard to GPU computation in other contexts
\cite{dziubak2012, broin2012,topi2012b}.
The above expansions assume that the spectrum of $H$ lies in the interval
$[-1, 1]$. For a Hamiltonian with spectrum beyond this range,
we need to shift and scale it, with a corresponding opposite scaling of the
time interval $\Delta t$.
The order of expansion $N_p$ depends on the time interval $\Delta t$ and
the desired accuracy.
The above summations can be efficiently evaluated
by using the following recursion relations ($m \geq 2$):
\begin{equation}
 T_m(H) = 2 H T_{m-1}(H) - T_{m-2}(H),
\end{equation}
\begin{equation}
[X, T_m(H)] = 2[X, H] T_{m-1}(H) + 2H [X, T_{m-1}(H)] - [X, T_{m-2}],
\end{equation}
\begin{equation}
 T_0(H) = 1 \quad
 T_1(H) = H,
\end{equation}
\begin{equation}
 [X, T_0(H)] = 0, \quad
 [X, T_1(H)] = [X,H].
\end{equation}

\subsection{GPU implementation}

In this subsection, we consider the GPU implementation of the algorithms.
We use CUDA \cite{cuda} as our developing tool. We only discuss the relevant
techniques of our CUDA implementation when appropriate; the reader is
referred to the official programming guide \cite{cuda} for more details.

To achieve high performance, we implement nearly all of the algorithms on
the GPU, minimizing data transfer between the CPU and the GPU.
Here we present the pseudo codes for calculating
$\rho (E)$, $\rho (E) C_{vv}(E, t)$, and $\rho (E) \Delta X^2(E, t)$
in Algorithms \ref{algorithm:DOS}, \ref{algorithm:DOS_times_VAC},
and \ref{algorithm:DOS_times_MSD}, respectively.
While for $\rho (E)$, we only need to calculate one set of moments,
for $\rho (E) C_{vv}(E, t)$ and $\rho (E) \Delta X^2(E, t)$, we have
to calculate a set of moments at each correlation time
$t_m$ $(0\leq m < N_c)$.
Thus, calculating
the conductivity is generally much more demanding
than calculating the DOS.
Note that we only calculate
the moments in the GPU, and copy their results
to the CPU for performing the Fourier transform
or the Chebyshev summation.
We could do all the calculations in the GPU, but
it does not result in a significant gain in
the overall performance,
since the calculation of the moments takes the majority
of the computation time.

In the previous subsection, we have written
$\rho (E) C_{vv}(E, t)$ and $\rho (E) \Delta X^2(E, t)$
in symmetric forms. The advantage is
that we can use the following iteration relations
to calculate the conductivity at different correlation times:
\begin{equation}
\label{equation:iteration_1}
 U^{\dagger}(t+\Delta t) V |\phi\rangle =
 U^{\dagger}(\Delta t) U^{\dagger}(t) V |\phi\rangle,
\end{equation}
\begin{equation}
\label{equation:iteration_2}
 \langle \phi| U(t+\Delta t) V   =
 \langle \phi| U(t) U(\Delta t) V,
\end{equation}
\begin{equation}
\label{equation:iteration_3}
[X, U(t+\Delta t)]  |\phi\rangle =
U(\Delta t) [X, U(t)]|\phi\rangle +
[X, U(\Delta t)] U(t) |\phi\rangle.
\end{equation}

\begin{algorithm}
\caption{Pseudo code for calculating $\rho (E)$.}
\label{algorithm:DOS}
\begin{algorithmic}[1]
\If {use the FTM}
    \State calculate $F_n^{\textmd{DOS}}$ in Eq. (\ref{equation:F_n_DOS}) in the GPU
    \State copy the $F_n^{\textmd{DOS}}$ data from the GPU to the CPU
    \State calculate $\rho(E)$ in the CPU using Eq. (\ref{equation:FTM_DOS})
\EndIf
\If {use the KPM}
    \State calculate $C_n^{\textmd{DOS}}$ in Eq. (\ref{equation:C_n_DOS}) in the GPU
    \State copy the $C_n^{\textmd{DOS}}$ data from the GPU to the CPU
    \State calculate $\rho(E)$ in the CPU using Eq. (\ref{equation:KPM_DOS})
\EndIf
\end{algorithmic}
\end{algorithm}

\begin{algorithm}
\caption{Pseudo code for calculating $\rho (E) C_{vv}(E, t)$.}
\label{algorithm:DOS_times_VAC}
\begin{algorithmic}[1]
\Require $|\phi\rangle$ is the initial random vector
\For {$m$ = 1 to $N_c -1$}
    \State calculate $U^{\dagger}(t_m)V|\phi\rangle$ iteratively
           using Eq. (\ref{equation:iteration_1})
    \State calculate $\langle \phi|U(t_m) V$ iteratively
           using Eq. (\ref{equation:iteration_2})
    \If {use the FTM}
      \State calculate $F_n^{\textmd{VAC}}$ in Eq. (\ref{equation:F_n_VAC}) in the GPU
      \State copy the $F_n^{\textmd{VAC}}$ data from the GPU to the CPU
      \State calculate $\rho(E)C_{vv}(E, t_m)$ in the CPU using Eq. (\ref{equation:FTM_VAC})
    \EndIf
    \If {use the KPM}
      \State calculate $C_n^{\textmd{VAC}}$ in Eq. (\ref{equation:C_n_VAC}) in the GPU
      \State copy the $C_n^{\textmd{VAC}}$ data from the GPU to the CPU
      \State calculate $\rho(E) C_{vv}(E, t_m)$ in the CPU using Eq. (\ref{equation:KPM_VAC})
    \EndIf
\EndFor
\end{algorithmic}
\end{algorithm}

\begin{algorithm}
\caption{Pseudo code for calculating $\rho (E) \Delta X^2(E, t)$.}
\label{algorithm:DOS_times_MSD}
\begin{algorithmic}[1]
\Require $|\phi\rangle$ is the initial random vector
\For {$m$ = 0 to $N_c -1$}
    \State calculate $[X, U(t_m)]|\phi\rangle$ iteratively
           using Eq. (\ref{equation:iteration_3})
    \If {use the FTM}
      \State calculate $F_n^{\textmd{MSD}}$ in Eq. (\ref{equation:F_n_MSD}) in the GPU
      \State copy the $F_n^{\textmd{MSD}}$ data from the GPU to the CPU
      \State calculate $\rho(E) \Delta X^2(E, t_m)$ in the CPU using Eq. (\ref{equation:FTM_MSD})
    \EndIf
    \If {use the KPM}
      \State calculate $C_n^{\textmd{MSD}}$ in Eq. (\ref{equation:C_n_MSD}) in the GPU
      \State copy the $C_n^{\textmd{MSD}}$ data from the GPU to the CPU
      \State calculate $\rho(E) \Delta X^2(E, t_m)$ in the CPU using Eq. (\ref{equation:KPM_MSD})
    \EndIf
\EndFor
\end{algorithmic}
\end{algorithm}

The calculation of the moments in both the FTM and the KPM
used in the above three algorithms can also be carried out iteratively.
We note that the Fourier moments in
equations (\ref{equation:F_n_DOS} - \ref{equation:F_n_MSD})
can be expressed in a unified way:
\begin{equation}
 F_n \approx \langle \phi_L| U(n\Delta t) |\phi_R\rangle.
\end{equation}
Different moments only differ in $|\phi_L\rangle$ and $|\phi_R\rangle$:
for DOS, $|\phi_L\rangle = |\phi_R\rangle = |\phi\rangle$;
for VAC, $|\phi_L\rangle = V U^{\dagger}(t)|\phi\rangle$ and
$|\phi_R\rangle = U^{\dagger}(t) V|\phi\rangle$;
for MSD, $|\phi_L\rangle = |\phi_R\rangle = [X, U(t)]|\phi\rangle$.
Similarly, the Chebyshev moments
in equations (\ref{equation:C_n_DOS} - \ref{equation:C_n_MSD})
can be expressed uniformly as
\begin{equation}
 C_n \approx \langle \phi_L| T_n(H) |\phi_R\rangle.
\end{equation}
Thus, we can present the calculations of these different moments
in a unified way, as shown in Algorithms \ref{algorithm:moments_FTM} and
\ref{algorithm:moments_KPM}.

\begin{algorithm}
\caption{Pseudo code for calculating the Fourier moments
$F_n = \langle \phi_L| U(n\Delta \tau) |\phi_R\rangle$.}
\label{algorithm:moments_FTM}
\begin{algorithmic}[1]
\State kernel: $|\phi'_R\rangle \leftarrow |\phi_R\rangle$
\State kernel: $F_0 \leftarrow \langle \phi_L |\phi'_R\rangle$
\For {$n$ = 1 to $N_m$}
   \State calculate $|\phi'_R\rangle \leftarrow U(\Delta \tau) |\phi'_R\rangle$ in the GPU
   \State kernel: $F_n \leftarrow \langle \phi_L |\phi'_R\rangle$
\EndFor
\State kernel: $|\phi'_R\rangle \leftarrow |\phi_R\rangle$
\For {$n$ = 1 to $N_m$}
   \State calculate $|\phi'_R\rangle \leftarrow U^{\dagger}(\Delta \tau) |\phi'_R\rangle$ in the GPU
   \State kernel: $F_{-n} \leftarrow \langle \phi_L |\phi'_R\rangle$
\EndFor
\end{algorithmic}
\end{algorithm}

\begin{algorithm}
\caption{Pseudo code for calculating the Chebyshev moments
$C_n = \langle \phi_L| T_n(H) |\phi_R\rangle$.}
\label{algorithm:moments_KPM}
\begin{algorithmic}[1]
\State kernel: $|\phi_0\rangle \leftarrow |\phi_R\rangle$
\State kernel: $C_0 \leftarrow \langle \phi_L |\phi_0\rangle$
\State kernel: $|\phi_1\rangle \leftarrow H |\phi_0\rangle$
\State kernel: $C_1 \leftarrow \langle \phi_L |\phi_1\rangle$
\For {$n$ = 2 to $N_m-1$}
   \State kernel: $|\phi_2\rangle \leftarrow 2H |\phi_1\rangle - |\phi_0\rangle$
   \State kernel: $C_n \leftarrow \langle \phi_L |\phi_2\rangle$
\EndFor
\end{algorithmic}
\end{algorithm}

We next consider the time-evolution of quantum states.
In Algorithms \ref{algorithm:evolve}
and \ref{algorithm:evolvex}, we present the algorithms
for evaluating $|\phi_{\textmd{out}}\rangle = U(\pm \Delta t)|\phi_{\textmd{in}}\rangle$
and $|\phi_{\textmd{out}}\rangle = [X, U(\Delta t)]|\phi_{\textmd{in}}\rangle$,
according to Eq. (\ref{equation:cheb_1}) and Eq. (\ref{equation:cheb_2}), respectively.
In Algorithm \ref{algorithm:evolve},
besides the input vector $|\phi_{\textmd{in}}\rangle$,
and the output vector  $|\phi_{\textmd{out}}\rangle$,
we need three auxiliary vectors, $|\phi_0\rangle$,
$|\phi_1\rangle$, and $|\phi_2\rangle$.
In Algorithm \ref{algorithm:evolvex}, we need another set of
auxiliary vectors, $|\phi_0^x\rangle$,
$|\phi_1^x\rangle$, and $|\phi_2^x\rangle$.
All of these vectors should be defined
in global memory in order to pass data between kernels.

\begin{algorithm}
\caption{Pseudo code for calculating
         $|\phi_{\textmd{out}}\rangle = U(\pm \Delta t)|\phi_{\textmd{in}}\rangle$}
\label{algorithm:evolve}
\begin{algorithmic}[1]
\State kernel: $|\phi_0\rangle \leftarrow |\phi_{\textmd{in}}\rangle$
\State kernel: $|\phi_1\rangle \leftarrow H |\phi_0\rangle$
\State kernel: $|\phi_{\textmd{out}}\rangle \leftarrow
               J_0\left(\frac{\Delta t}{\hbar}\right) |\phi_0\rangle
               + 2(\mp i)J_1\left(\frac{\Delta t}{\hbar}\right) |\phi_1\rangle$
\For {$m$ = 2 to $N_p -1$}
    \State kernel: $|\phi_2\rangle \leftarrow 2 H |\phi_1\rangle - |\phi_0\rangle$
    \State kernel: $|\phi_{\textmd{out}}\rangle
                   \leftarrow |\phi_{\textmd{out}}\rangle
                   + 2(\mp i)^m J_m\left(\frac{\Delta t}{\hbar}\right) |\phi_2\rangle$
    \State Permute pointers: $|\phi_0\rangle \leftarrow |\phi_1\rangle$,
                             $|\phi_1\rangle \leftarrow |\phi_2\rangle$,
                             $|\phi_2\rangle \leftarrow |\phi_0\rangle$
\EndFor
\end{algorithmic}
\end{algorithm}

\begin{algorithm}
\caption{Pseudo code for calculating
         $|\phi_{\textmd{out}}\rangle = [X,U(\Delta t)]|\phi_{\textmd{in}}\rangle$}
\label{algorithm:evolvex}
\begin{algorithmic}[1]
\State kernel: $|\phi_0\rangle \leftarrow |\phi_{\textmd{in}}\rangle$
\State kernel: $|\phi_0^x\rangle \leftarrow 0$
\State kernel: $|\phi_1\rangle \leftarrow H |\phi_0\rangle$
\State kernel: $|\phi_1^x\rangle \leftarrow [X, H] |\phi_{\textmd{in}}\rangle$
\State kernel: $|\phi_{\textmd{out}}\rangle
               \leftarrow 2(-i)J_1\left(\frac{\Delta t}{\hbar}\right) |\phi_1^x\rangle$
\For {$m$ = 2 to $N_p -1$}
    \State kernel: $|\phi_2\rangle \leftarrow 2 H |\phi_1\rangle - |\phi_0\rangle$
    \State kernel: $|\phi_2^x\rangle \leftarrow 2 [X, H] |\phi_1\rangle
                   + 2 H |\phi_1^x\rangle - |\phi_0^x\rangle$
    \State kernel: $|\phi_{\textmd{out}}\rangle
                   \leftarrow |\phi_{\textmd{out}}\rangle + 2(-i)^m
                   J_m\left(\frac{\Delta t}{\hbar}\right) |\phi_2^x\rangle$
    \State Permute pointers: $|\phi_0\rangle \leftarrow |\phi_1\rangle$,
                             $|\phi_1\rangle \leftarrow |\phi_2\rangle$,
                             $|\phi_2\rangle \leftarrow |\phi_0\rangle$
    \State Permute pointers: $|\phi_0^x\rangle \leftarrow |\phi_1^x\rangle$,
                             $|\phi_1^x\rangle \leftarrow |\phi_2^x\rangle$,
                             $|\phi_2^x\rangle \leftarrow |\phi_0^x\rangle$
\EndFor
\end{algorithmic}
\end{algorithm}

An examination of Algorithms \ref{algorithm:evolve}
and \ref{algorithm:evolvex} reveals that,
apart from some simple linear transformations,
the only nontrivial calculations are the matrix-vector multiplications,
$|\phi_{\textmd{out}}\rangle = H |\phi_{\textmd{in}}\rangle$
and $|\phi_{\textmd{out}}\rangle = [X, H]|\phi_{\textmd{in}}\rangle$.
In Algorithm \ref{algorithm:Apply_Hamiltonian},
we present the pseudo code of the CUDA kernel which evaluates
$|\phi_{\textmd{out}}\rangle = H |\phi_{\textmd{in}}\rangle$;
the evaluation of
$|\phi_{\textmd{out}}\rangle = [X, H]|\phi_{\textmd{in}}\rangle$
is very similar.

The strategy in Algorithm \ref{algorithm:Apply_Hamiltonian}
is to use one thread for one element of the output vector.
By using a block size of $S_b$, the number of blocks in the kernel
is $(N-1)/S_b+1$, where $N$ is the number of sites in the system.
Thus, this kernel is executed with the
configuration of $<<<(N-1)/S_b+1, S_b>>>$.
The \textbf{if} statement on line 1 is necessary
to avoid manipulating invalid memory
in the case of $N$ not being an integer multiple of $S_b$.
Lines 2-7 are devoted to the calculation of
$\phi_{\textmd{out}}[n]$,
where the variable temp is used to reduce the global memory access,
which is very time-consuming. We use a neighbor list to
specify the Hamiltonian, denoting the number of neighbors
to site $n$ as NN$_n$,
and indexing the $k$th neighbor of site $n$
as NL$_{nk}$. For a sparse Hamiltonian, NN$_n$
is much smaller than the total number of sites $N$.
The NL$_{nk}$ data should be coded in such a way that
the indices of the $k$th neighbor sites for all the
sites are stored consecutively, i.e., in the order of
NL$_{00}$, NL$_{10}$, NL$_{20}$, $\cdots$,
NL$_{01}$, NL$_{11}$, NL$_{21}$, $\cdots$,
NL$_{0k}$, NL$_{1k}$, NL$_{2k}$, $\cdots$.
This special order ensures coalescing in global memory access,
which means that consecutive threads access consecutive data
in the global memory. This requirement has also been noticed
in our previous work on molecular dynamics simulations \cite{fan2013}.

\begin{algorithm}
\caption{The algorithm for evaluating
         $|\phi_{\textmd{out}}\rangle = H|\phi_{\textmd{in}}\rangle$}
\label{algorithm:Apply_Hamiltonian}
\begin{algorithmic}[1]
\Require $\phi_{\textmd{in}}[m]$ is the $m$th component of
         $|\phi_{\textmd{in}}\rangle$
\Require $\phi_{\textmd{out}}[n]$ is the $n$th component of
         $|\phi_{\textmd{out}}\rangle$
\Require $n$ = blockIdx.x * blockDim.x + threadIdx.x
\Require $N$ is the number of sites in the system
\Require NN$_n$ is the total number of neighbor sites of site $n$
\Require NL$_{nk}$ is the index of the $k$th neighbor site of site $n$
\If {$n < N$}
    \State temp $\leftarrow 0$
    \For {$k$ = 0 to NN$_n - 1$}
        \State $m \leftarrow \textmd{NL}_{nk}$ 
        \State temp $\leftarrow$ temp + $H_{nm}
                \phi_{\textmd{in}}[m] $
    \EndFor
        \State $\phi_{\textmd{out}}[n] \leftarrow$ temp
\EndIf
\end{algorithmic}
\end{algorithm}

\section{Performance evaluation}
\label{section:Performance}

In this section, we compare the relative performance of our GPU
and CPU implementations, and the relative performance of the
FTM and the KPM.

\subsection{GPU versus CPU}

We firstly evaluate the relative performance of
our GPU implementation with respect to our CPU implementation.
The comparison is made between a Tesla
K20 GPU card and an Intel Xeon E5-1620 @ 3.60 GHz CPU core.
The serial CPU code is implemented
in C/C++ and is compiled with an O3 optimization mode. Although the algorithms
in the previous section are presented by using a complex number notation,
in both the CPU and the GPU implementation, we use two real vectors
for a complex state vector, which can save nearly half of the calculations
compared with a naive use of the intrinsic complex number. Both the CPU and
the GPU code use double-precision arithmetics.

The major computation which scales linearly with the system size is
the Chebyshev iteration, which is used for both the time-evolution
and the KPM. We thus present a performance evaluation of the Chebyshev
iteration part of the code in some detail.
We chose to present the testing results for
$|\phi_2\rangle = 2 H |\phi_1\rangle - |\phi_0\rangle$;
those for
$|\phi_2^x\rangle = 2 [X, H] |\phi_1\rangle
+ 2 H |\phi_1^x\rangle - |\phi_0^x\rangle$
are similar.

\begin{figure}
\begin{center}
  \includegraphics[width=3 in]{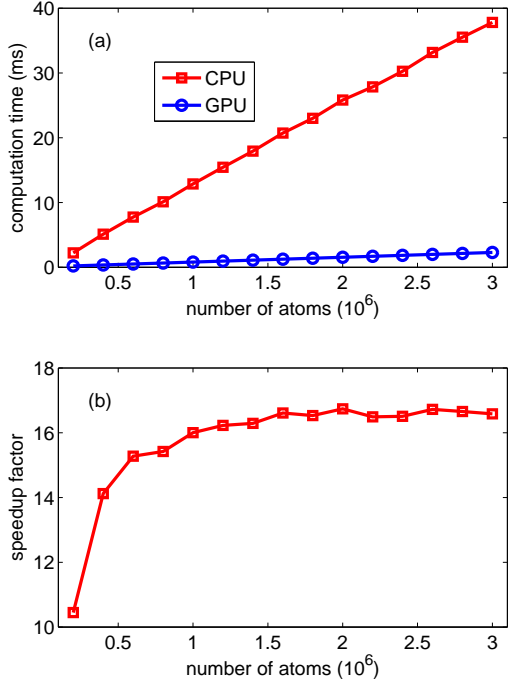}
  \caption{(Color online) (a) Computation times for evaluating
           $|\phi_2\rangle = 2 H |\phi_1\rangle - |\phi_0\rangle$
           in the CPU and the GPU and (b) the corresponding speedup factor
           as a function of the system size. Double-precision is used
           for both the CPU and the GPU code.}
  \label{figure:cpu_vs_gpu}
\end{center}
\end{figure}

Figure \ref{figure:cpu_vs_gpu} shows the results of the performance
evaluation of the Chebyshev iteration part, where the speedup factor is defined
as the computation time in the CPU over that in the GPU. The computational time
in the CPU scales linearly with respect to the simulation size, which reflects
the linear-scaling nature of the algorithm.
The computation time in the GPU also scales linearly approximately.
The speedup factor increases from about 10.5 to about 16.5 with the number of atoms
in the simulated system increasing from 0.2 million to 1.6 million and nearly
saturates thereafter. For all the other calculations
such as the evaluation of the inner products, we also obtained a
comparable speedup factor. The overall speedup factor
of our GPU implementation
over our CPU implementation is observed to be about 16.

This speedup factor seems to be not very impressive.
Indeed, in our recent work on exact
diagonalization of the Hubbard model using the LRM
on the GPU \cite{topi2012a}, a speedup factor of about 60 is obtained
using double-precision. The difference in the speedup factor
results from the different computational intensities of the
problems. For example, in the Hubbard model,
for a Hamiltonian size of 853776 (12 spin sites),
the computation times for one Lanczos iteration in the CPU
and the GPU  are about 120 ms and 2 ms, respectively,
giving a speedup factor of 60 \cite{topi2012a}.
In comparison, for our tight-biding model with a Hamiltonian size of
$10^6$, the computation times for one Chebyshev iteration in the
CPU and the GPU are about 12.8 ms and 0.8 ms, giving a speedup factor of 16.
We see that for a given Hamiltonian size, the Hubbard model is
about 10 times more computationally intensive than the single-particle
tight-binding model and attains a higher speedup factor. Similar
dependence of the speedup factor on the computational intensity
has also been observed in our recent work on molecular dynamics
simulation \cite{fan2013}.

\subsection{KPM versus FTM}

\begin{figure}
\begin{center}
  \includegraphics[width=3 in]{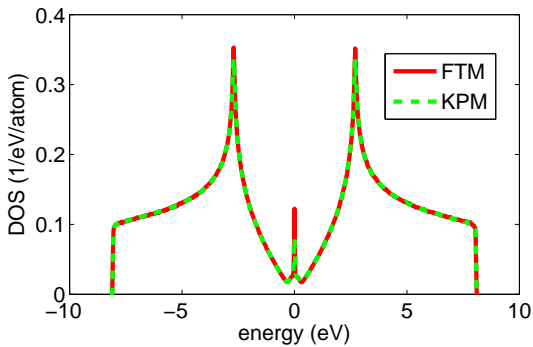}
  \caption{(Color online) DOS as a function of energy for
           2D  graphene of size $2048\times 1024$
           (using 10 random vectors)
           with defect density $n=1\%$
           calculated by the KPM with $N_m = 1000$ and the FTM
           with $2 N_m + 1= 1001$.
           }
  \label{figure:kpm_vs_ftm}
\end{center}
\end{figure}

We then give a comparison of the relative performance of the KPM and the FTM.
For the FTM, the calculation of each Fourier moment involves a time-evolution
with a time step $\Delta \tau$. The choice of the time step used in the FTM
is related to the Nyquist sampling rates used in digital signal analysis:
it should not be too large to give aliasing errors, and not too small to
reduce the energy resolution \cite{feit1982}.
The optimal value of $\Delta \tau$ corresponding
to a maximum bandwidth $\Delta E$ of the energy spectrum without
aliasing error can be fixed to be
\begin{equation}
 \Delta \tau = \frac{2 \pi \hbar}  {\Delta E}.
\end{equation}
For a scaled Hamiltonian with spectrum $[-1, 1]$, we have  $\Delta E = 2$
and $\Delta \tau = \pi \hbar$. Then, the dimensionless argument in
the Bessel function is $\pi$, which determines the number of
Chebyshev iterations in the time evolution operator $U(\Delta \tau)$
to be about $N_p = 20$ for an accuracy of $10^{-15}$. In contrast, the
calculation of each Chebyshev moment in the KPM only involves one
Chebyshev iteration.

To give a fair comparison of the relative efficiency, we should also
consider the energy resolution $\delta E$, which is related to
the number of moments $2N_m+1$ in the FTM and $N_m$ in the KPM.
Quantitatively, we have
\begin{equation}
 \delta E = \frac{2 \pi \hbar}{\Delta\tau (2 N_m + 1)} = \frac{\Delta E}{2 N_m + 1}
\end{equation}
in the FTM \cite{feit1982} and
\begin{equation}
 \delta E = \frac{\pi \Delta E}{N_m}
\end{equation}
in the KPM \cite{weibe2006}, respectively.
Figure \ref{figure:kpm_vs_ftm} gives a comparison
of the DOSs calculated by the
the KPM with $N_m = 1000$ and the FTM
with $2 N_m + 1= 1001$. We see that they give
consistent results and the FTM indeed has a higher
energy resolution when using the same number of moments.

By combining the above analysis, we come to the conclusion that
the KPM is about $20/\pi \approx 6.4$ times as efficient as the FTM for
achieving the same energy resolution. However,
for the transport simulations, this
difference of efficiency only matters in the diffusive
regime, where the correlation time step $\Delta t$ should
be relatively small, and the computation time is dominated
by the calculation of the $\delta$ function. In the localized
regime, where the correlation time step is usually chosen to be
very large, the computation time is dominated by the time-evolution
$[X, U(t)]|\phi\rangle$, and the relative efficiency of the KPM over the
FTM does not lead to a significant gain in performance for the
whole simulation.

\section{Validation}
\label{section:validation}

In this section, we validate our GPU code by studying the transport
properties of 2D graphene and quasi-1D graphene nanoribbons
in both the ballistic, the diffusive and the localized regimes.

\subsection{The ballistic transport regime}

\begin{figure}
\begin{center}
  \includegraphics[width=3 in]{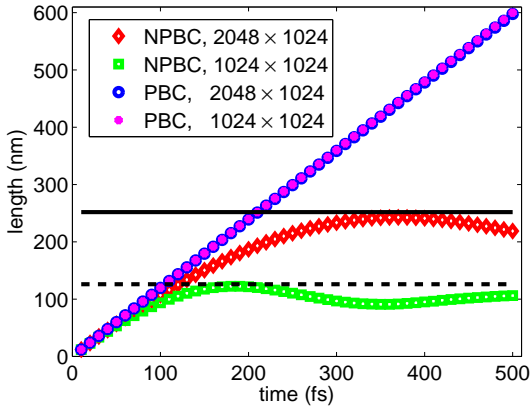}
  \caption{(Color online) Length defined by $L = 2 \sqrt{\Delta X^2(E, t)}$
           as a function of the correlation time for pristine graphene
           with different
           sizes and different boundary conditions along the transport direction
           (the zigzag direction):
           periodic (PBC) and non-periodic (NPBC).
           The solid and dashed horizontal lines indicate the
           sample lengths along the transport direction:
           126 nm and 252 nm for $1024\times 1024$ graphene
           and $2048\times 1024$ graphene, respectively.}
  \label{figure:pbc_vs_npbc}
\end{center}
\end{figure}

\begin{figure*}
\begin{center}
  \includegraphics[width=3 in]{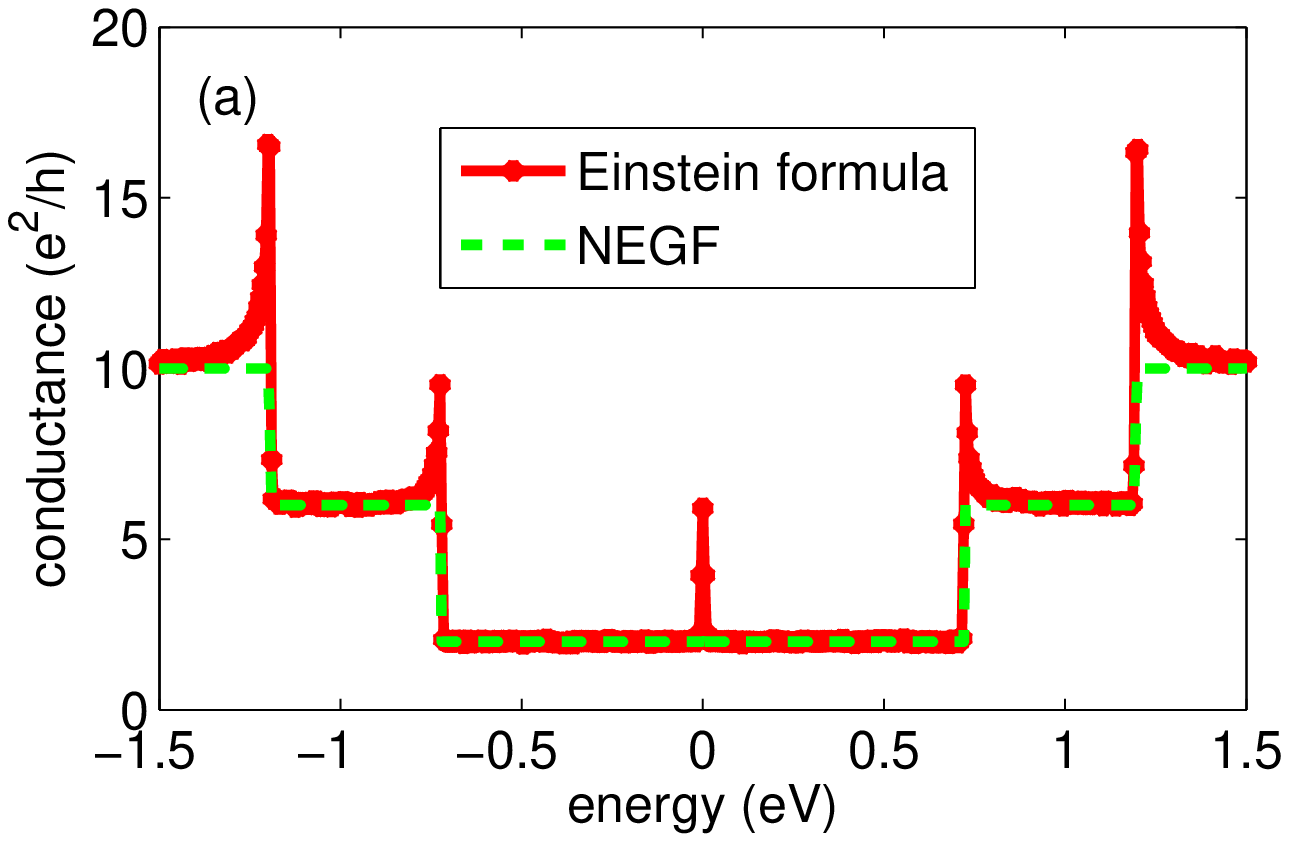}
  \includegraphics[width=3 in]{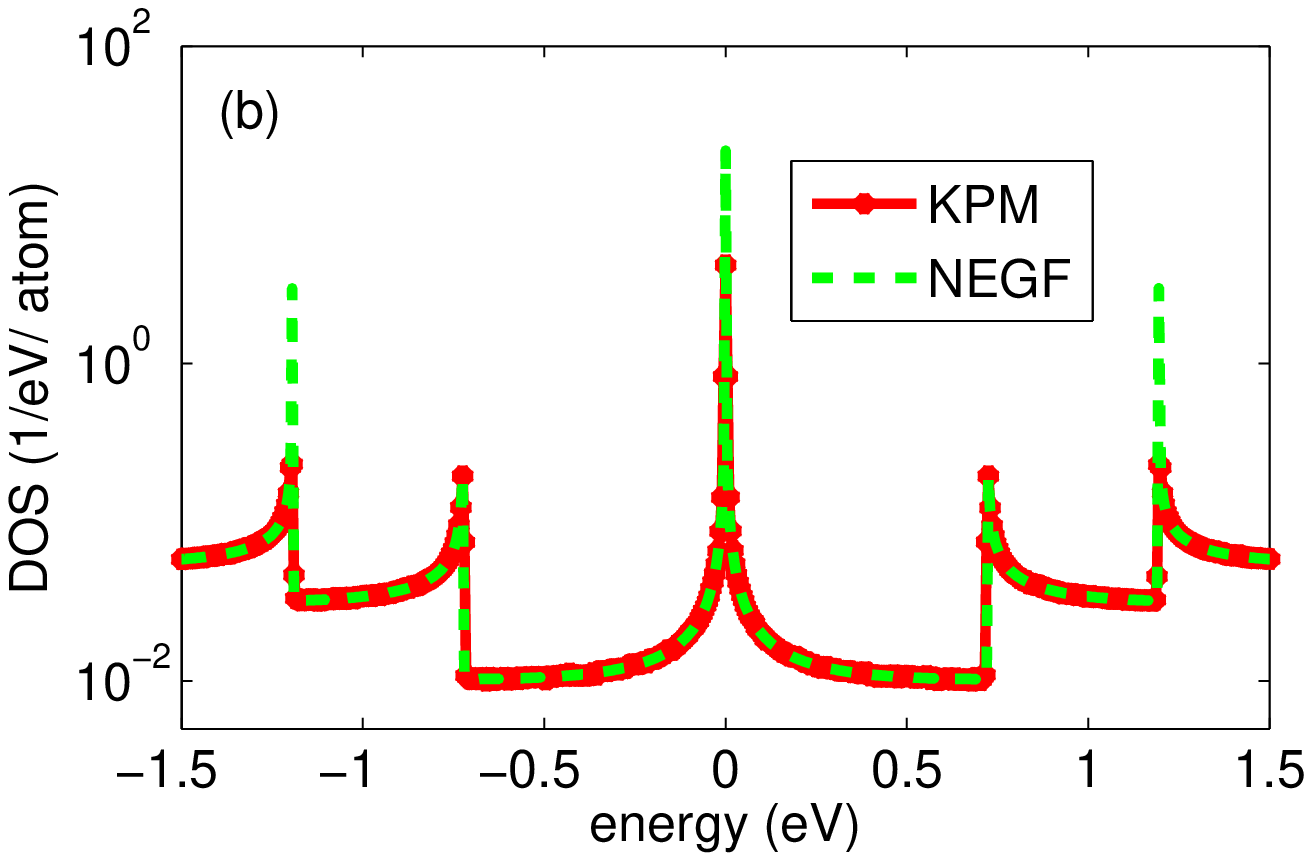}\\
  \includegraphics[width=3 in]{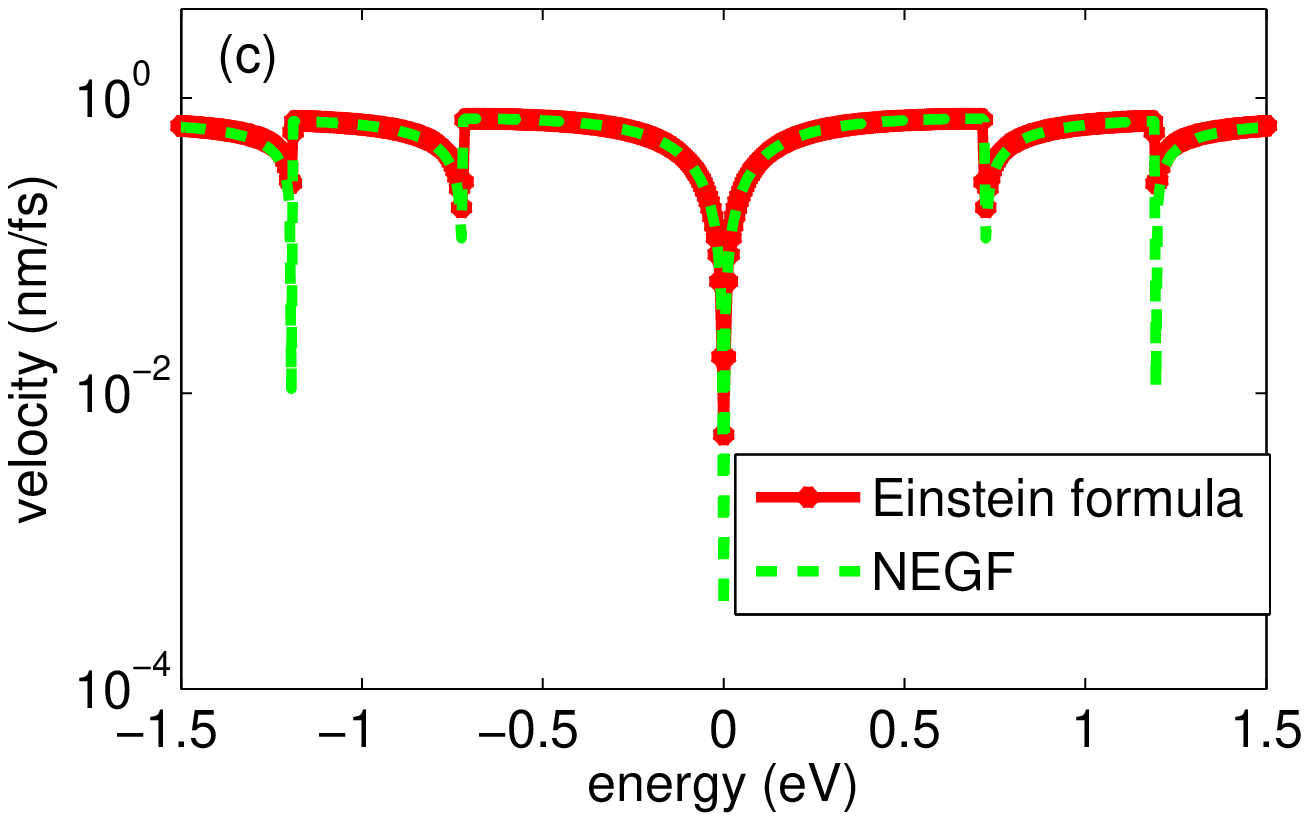}
  \includegraphics[width=3 in]{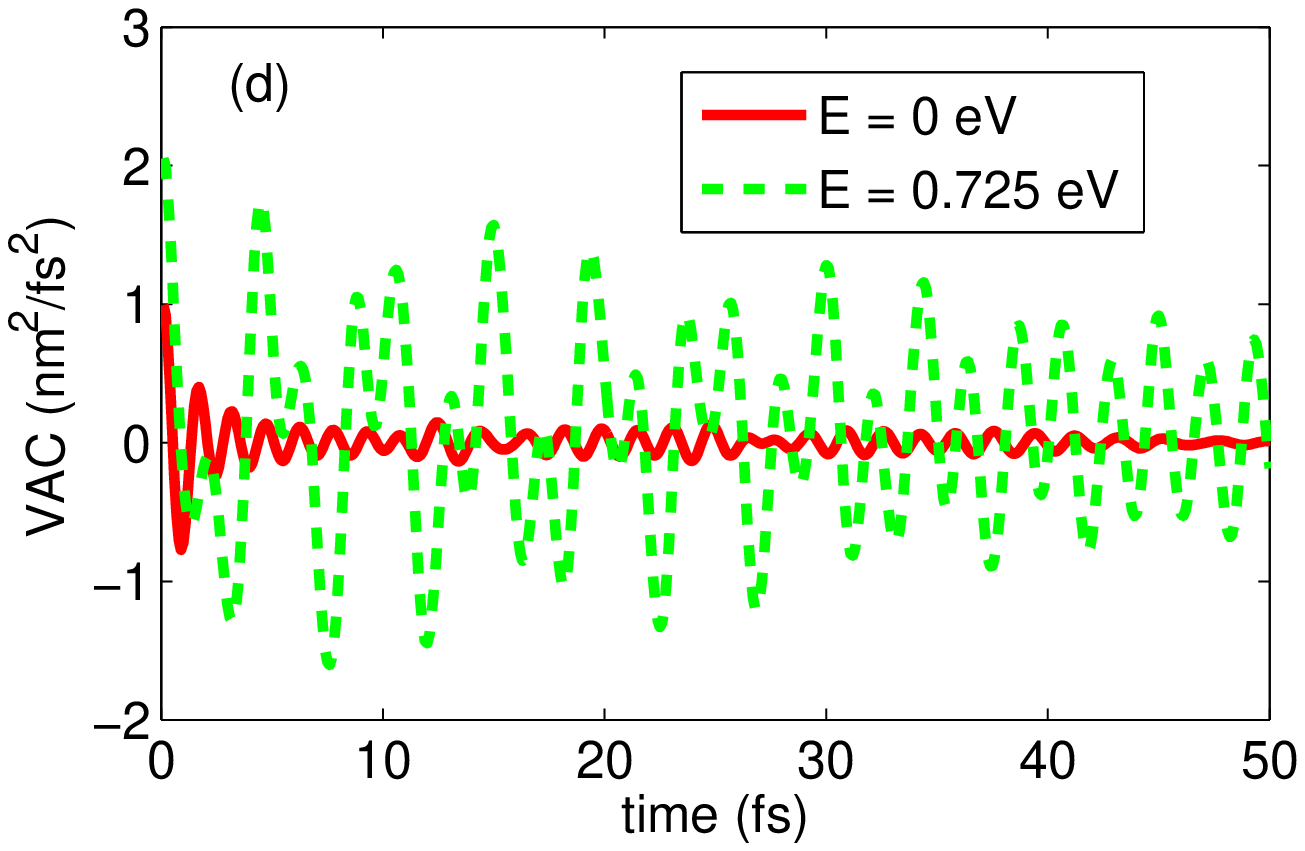}
  \caption{(Color online)
            Ballistic transport properties for pristine ZGNR of size
            $65536 \times 16$ (using 100 random vectors).
            (a) Ballistic conductance as a function of energy
            calculated by the Einstein formula
            (Eq. (\ref{equation:ballistic_conductance}))
            and the NEGF method. (b) DOS calculated by the KPM and the NEGF
            method. (c) $\sqrt{v^2(E)}$ as a function of energy
            deduced from the calculated
            DOS and conductance by using
            Eq. (\ref{equation:ballistic_conductance_v}). (d) VACs
            as a function of correlation time
            for the Dirac point and the next band edge.  }
  \label{figure:ballistic}
\end{center}
\end{figure*}

For ballistic transport without any scattering, the VAC does not decay
with time, resulting in a divergent conductivity.
A finite conductance can only be deduced by introducing a length scale.
While there is no intrinsic definition of length in the
Green-Kubo and the Einstein formulas, a definition of length
in terms of the MSD,
\begin{equation}
\label{equation:length}
 L(E, t) = 2 \sqrt{\Delta X^2(E, t)}
\end{equation}
is frequently used \cite{triozon2002,leconte2011,lherbier2012}.
The conductance of a system with width $W$ can be defined as
\begin{equation}
\label{equation:ballistic_conductance}
 G(E) = \frac{W}{L(E, t)} \sigma^{\textmd{E1}}(E, t).
\end{equation}
Although the correlation time $t$ appears in the above equation,
a converged time-independent (length-independent) value of $G(E)$
can be obtained within a short correlation time.
We note that the factor of 2 in the above length definition is necessary
to obtain correct results if we use the correct definition
of conductivity, $\sigma^{\textmd{E1}}(E, t)$,
rather than the alternative, $\sigma^{\textmd{E2}}(E, t)$,
which is half of $\sigma^{\textmd{E1}}(E, t)$ in the ballistic regime.

To justify the factor of 2 in Eq. (\ref{equation:length}),
we examine the time-dependence of the length for pristine graphene
with different sizes and different boundary conditions
along the transport direction, which is chosen to be the zigzag
direction. By applying periodic boundary conditions
in the transport direction, there is no noticeable
difference in the results obtained by using
a longer sample (252 nm for $2048\times 1024$ graphene)
and a shorter sample (126 nm for $1024\times 1024$ graphene),
which reflects the small finite size effect in
Green-Kubo-like formulas \cite{fan2013}. In contrast, by imposing
a non-periodic boundary condition in the transport direction,
the diffusion of electrons is confined by the sample size, with
the maximum diffusion length as defined in Eq. (\ref{equation:length})
being the length of the sample. The factor of 2 can also be understood
intuitively: $\sqrt{\Delta X^2(E, t)}$ is the absolute diffusion
distance in one direction, and the factor of 2 accounts for the
diffusion in the opposite direction.

We now study the ballistic transport properties of a $65536\times 16$
pure ZGNR by comparing the results with those obtained by the
NEGF method. As can be seen from
Fig. \ref{figure:ballistic} (a), the overall plateaus
of the quantized conductance can be correctly produced by
Eq. (\ref{equation:ballistic_conductance}),
but the conductances around the band edges are overestimated.
Markussen \textit{et al}. \cite{markussen2006} also noticed this
problem and argued that the overshoots near the band edges
originate from the nonequivalence between the expectation values
of $v(E)$ and the square root of the expectation value of $v^2(E)$.
Here, we give an analysis of this problem from the numerical perspective.

In the ballistic regime, the VAC oscillates around some value (see
Fig. \ref{figure:ballistic} (d) for an example), and an average
value of $v^2(E)$ can be well established over a short correlation time.
Thus we can express the MSD as $\Delta X^2(E, t) = v^2(E) t^2$,
which results in the following expression for the conductance:
\begin{equation}
\label{equation:ballistic_conductance_v}
 G(E) = \frac{W}{2} e^2 \rho(E) \sqrt{v^2(E)}.
\end{equation}
Fig. \ref{figure:ballistic} (b) presents the calculated DOS and
Fig. \ref{figure:ballistic} (c) presents the deduced $\sqrt{v^2(E)}$.
We see that both $\rho(E)$ and $\sqrt{v^2(E)}$ are singular
near the band edges. Thus, the calculation of ballistic conductance
in the Einstein formalism involves multiplications of big
and small numbers, which is numerically unstable. Since the MSD
and the VAC are squared quantities, we obtain
an overestimation rather than an underestimation of the conductance.

\subsection{The diffusive transport regime}

\begin{figure*}
\begin{center}
  \includegraphics[width=2.2 in]{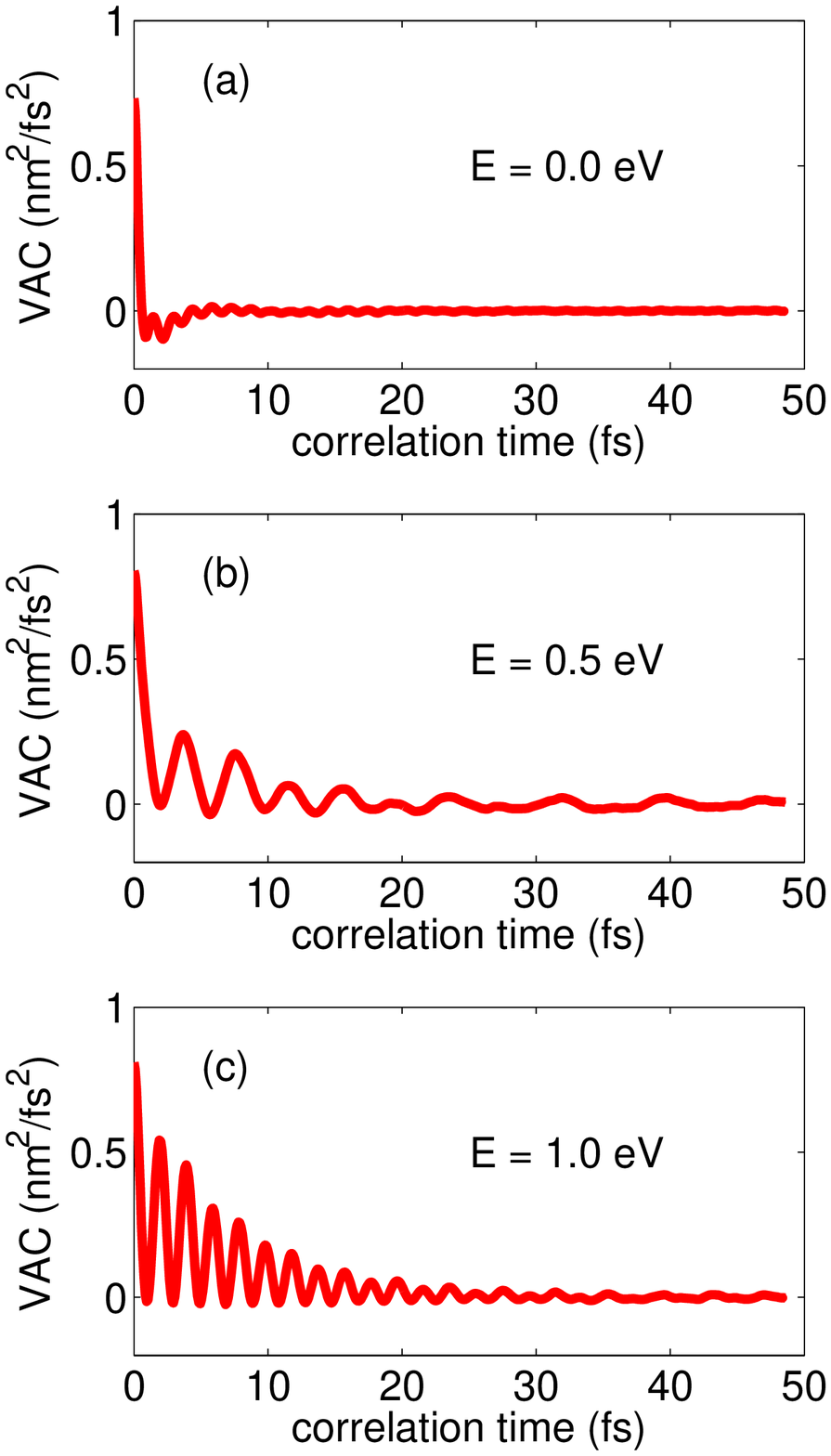}
  \includegraphics[width=2.2 in]{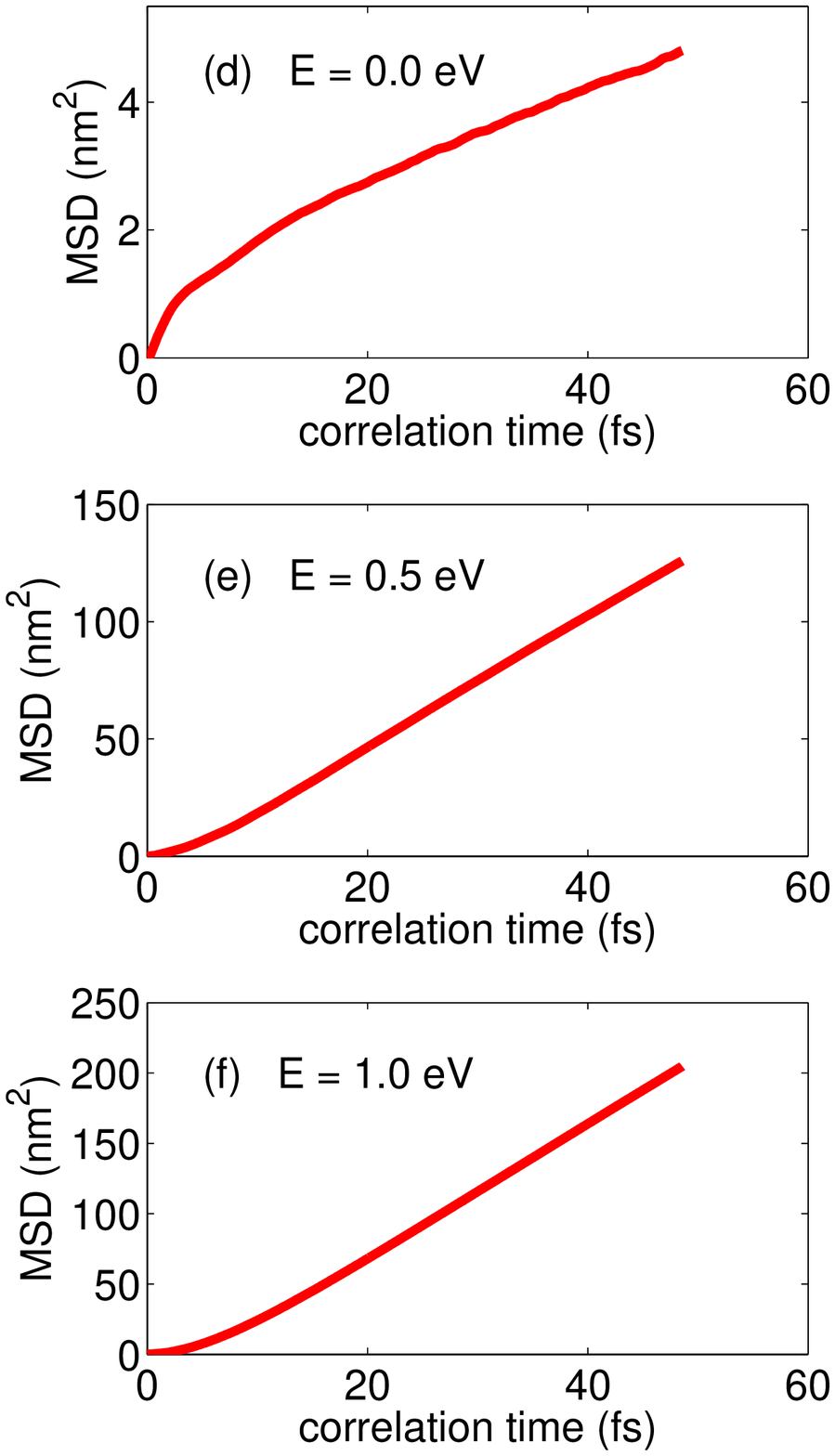}
  \includegraphics[width=2.2 in]{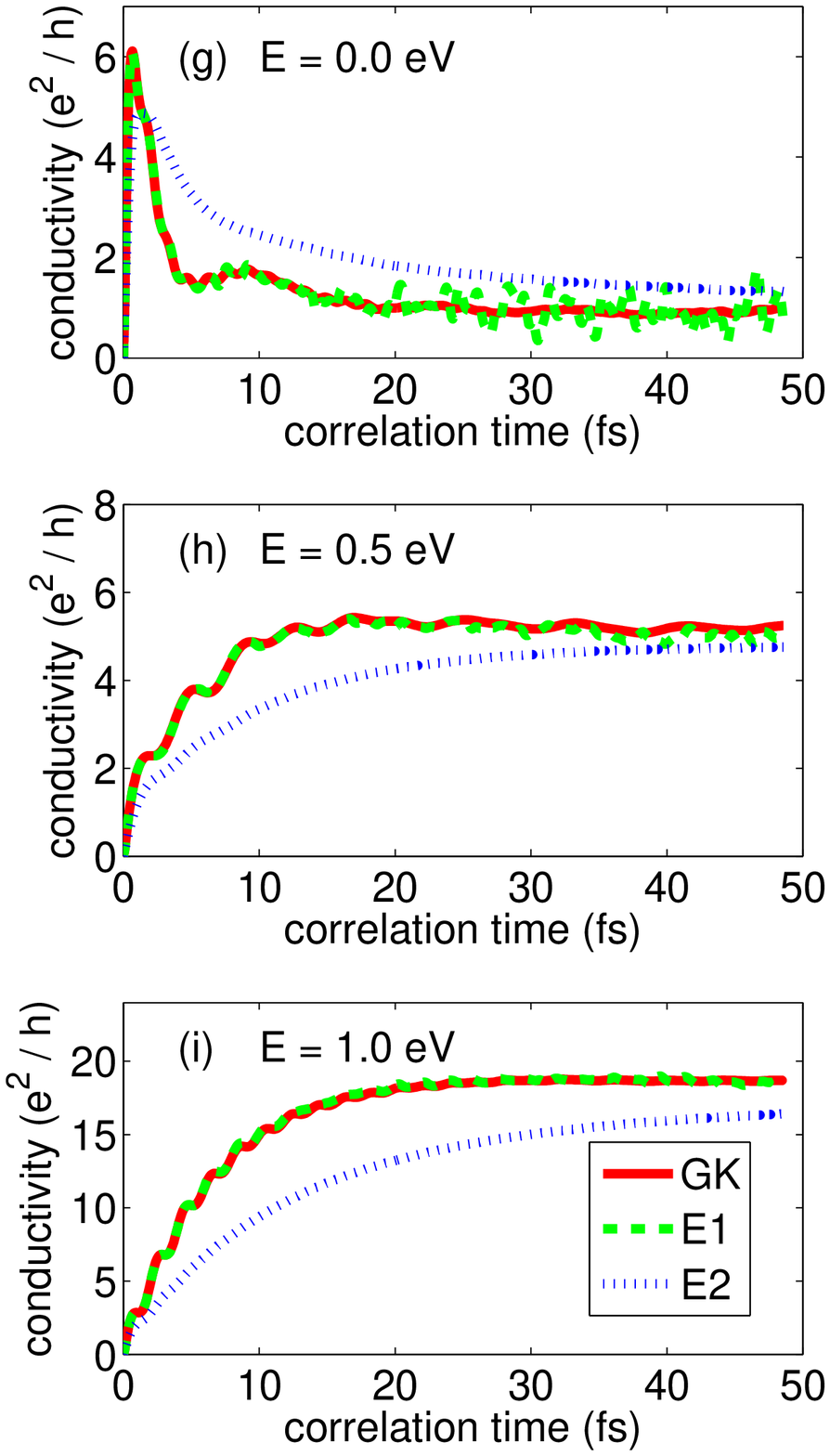}
  \caption{(Color online) (a-c) VACs, (d-f) MSDs,
           and (g-i) RECs as a function of correlation time
           at different energies for 2D graphene
           of size $2048 \times 1024$ (using 30 random vectors)
           with defect concentration $n=1\%$
           calculated by the KPM with $N_m = 3000$.
           For the RECs, the legends ``GK'', ``E1'', and ``E2'' refer to
           Eq. \ref{equation:REC_GK},
           Eq. \ref{equation:REC_E1}, and
           Eq. \ref{equation:REC_E2}, respectively.
           }
  \label{figure:diffusive}
\end{center}
\end{figure*}

We now turn to discuss the diffusive transport regime. We consider 2D graphene
of size $2048 \times 1024$ with defect concentration $n=1\%$. We use
both the Green-Kubo formula and the Einstein formula. The time
step is chosen to be $\Delta t = 0.1$ fs, small enough to
exhibit the detailed features of the ballistic-to-diffusive transition.

Figure \ref{figure:diffusive} (a-c) shows
the VACs for different energies as a function of correlation time.
We see that the VAC does not decay monotonically.
For the Dirac point $E=0.0$ eV, the VAC decays to zero within
one fs and then develops negative values up to 5 fs, after which
the VAC stays at zero for a relatively long time.
For higher energies, $E=0.5$ eV and 1.0 eV, apart from the expected
exponential decay, there is also an oscillatory component.
This oscillation has been discussed by de Laissardiere \textit{el al.}
\cite{laissardiere2011}, and is attributed to
the Zitterbewegung effect. A spectral analysis shows that the frequency
$\omega$ of the oscillation is directly related to the electron energy
by $\omega = 2 E/\hbar$, which is consistent with the
oscillation factor $\cos^2(Et/\hbar)$ in
the VAC \cite{laissardiere2011}.
By going from the Green-Kubo to the Einstein formalism,
these oscillations are smoothed out, as shown by the MSD curves in
Figure \ref{figure:diffusive} (d-f).
The ballistic-to-diffusive transition
is featured by the decay of the VAC in the Green-Kubo formalism,
or the quadratic-to-linear transition of the MSD in the Einstein
formalism.

From the VAC and the MSD, we can calculate RECs,
$\sigma^{\textmd{GK}}(E, t)$, $\sigma^{\textmd{E1}}(E, t)$, and
$\sigma^{\textmd{E2}}(E, t)$, as shown in
Fig. \ref{figure:diffusive} (g-i).
We see that the derivative-based definition of the REC in the
Einstein formalism is equivalent
to the REC defined in the Green-Kubo formalism:
$\sigma^{\textmd{E1}}(E, t) = \sigma^{\textmd{GK}}(E, t)$.
In contrast, the division-based definition of the REC in the
Einstein formalism deviates from the other two
in the ballistic-to-diffusive regime. One may note that
for the Dirac point, $\sigma^{\textmd{E1}}(E, t)$
has large fluctuations when $t>10$ fs. This reflects the
numerical difficulty of calculating the derivative
in Eq. (\ref{equation:REC_E1}),
especially for small time steps, and is probably the reason
for the preference of using $\sigma^{\textmd{E2}}(E, t)$
instead of $\sigma^{\textmd{E1}}(E, t)$ in some previous works.
However, we stress
that $\sigma^{\textmd{E2}}(E, t)$ is a wrong definition
in principle and should be used with caution.

\begin{figure}
\begin{center}
  \includegraphics[width=3 in]{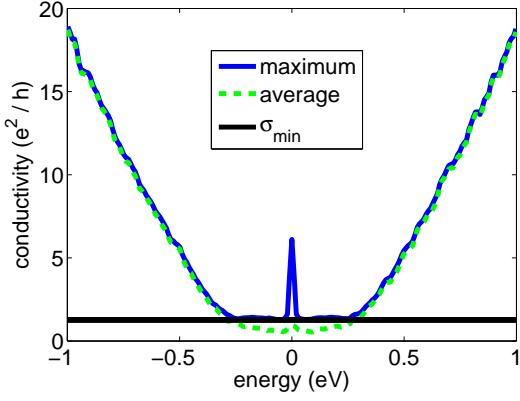}
  \caption{(Color online) Semi-classical conductivity
   of 2D graphene of size $2048 \times 1024$ (using 30 random vectors)
   with defect concentration $n=1\%$ calculated by 
   Eq. (\ref{equation:sigma_sc_maximum}) (labeled by
   ``maximum'') and Eq. (\ref{equation:sigma_sc_average})
   (labeled by ``average''), using $\sigma^{\textmd{GK}}(E,t)$. The horizontal line denotes the 
   value of $\sigma_{\textmd{min}} = 4e^2/(\pi h)$.
           }
  \label{figure:diffusive_sigma}
\end{center}
\end{figure}

The most interesting quantity in the diffusive regime is the
semi-classical conductivity, $\sigma_{\textmd{sc}}$,
which is conventionally
defined \cite{triozon2004,markussen2006,ishii2010,lherbier2008a,lherbier2008b,
laissardiere2011,leconte2011,radchenko2012,
lherbier2012,tuan2013,cresti2013} to
be the maximum value of the REC:
\begin{equation}
\label{equation:sigma_sc_maximum}
\sigma_{sc}(E) = \max\{\sigma(E, t), t>0\}.
\end{equation}
Using this definition, the calculated $\sigma_{\textmd{sc}}(E)$ 
(the solid line in Fig. \ref{figure:diffusive_sigma})
exhibits a plateau of minimum conductivity $\sigma_{\textmd{min}}=4e^2/(\pi h)$
in the range of $|E| <$ 0.25 eV, along
with a peak around the Dirac point.
Similar results have been obtained by Yuan \textit{et al.}
using the Green-Kubo formula
\cite{yuan2010} and by Cresti \textit{et al.}
using the Einstein formula \cite{cresti2013}.
One may note that the peaks found by Yuan \textit{et al.} \cite{yuan2010} 
are much lower than those found by Cresti \textit{et al.} \cite{cresti2013}. 
This difference
partly results from the different numerical approaches, but the
major reason is that Cresti \textit{et al.} use
Eq. (\ref{equation:sigma_sc_maximum}) to calculate 
$\sigma_{\textmd{sc}}(E)$, while Yuan \textit{et al.}
just integrate the VAC to some given correlation time.

A comment on the connection and difference between 
the Green-Kubo method in our work and the numerical approach
developed by Yuan \textit{et al.} is in order. After some algebra,
we can rewrite their formula for DC conductivity (Eq. (41) in
Ref. \cite{yuan2010}) using our notations as:
\begin{equation}
\label{equation:yuan2010}
 \sigma(E, t)  \approx \frac{2 e^2}{\Omega} \int_0^t \textmd{Re} 
              \left[ \langle \phi| 
               V U^{\dagger}(t) V U(t)
               | \delta(E-H) |\phi\rangle\right],
\end{equation}
which is equivalent to Eq. (\ref{equation:REC_GK}) 
and Eq. (\ref{equation:trace_VAC}) in our work. 
The difference between our
approach and their is mainly related to the numerical 
implementations. They firstly precompute all the
``quasi-eigenstates'' $|E_m\rangle = \delta(E_m-H)|\phi\rangle
\approx \frac{\Delta \tau}{2 \pi \hbar} 
\sum_{n=-N_m}^{+N_m} e^{i E_m n \Delta \tau/ \hbar}
U(n\Delta \tau) |\phi\rangle$ for a given number of
energy points $E_m$ and then store them in memory, before
calculating  $\sigma(E, t)$ using Eq. (\ref{equation:yuan2010}).
This strategy may be very efficient, but is not
economic in terms of memory usage, restricting the number of energy
points considered in one simulation to be around 64 \cite{yuan2010b}.

Although Eq. (\ref{equation:sigma_sc_maximum}) has been
widely used, there is no rigorous justification for
using it. The reason for choosing this definition may be related to
the unavoidable localization effects 
\cite{laissardiere2011,leconte2011,lherbier2012,radchenko2012,tuan2013,cresti2013}
in most of the
problems studied by this method. When localization takes
place, the REC decays with increasing correlation time
after achieving the diffusive regime, and it
is difficult to apply Eq. (\ref{equation:limit})
to find a time-independent (length-independent)
$\sigma_{\textmd{sc}}(E)$. Although 
Eq. (\ref{equation:sigma_sc_maximum})
works fine for higher energies, it is problematic
near the Dirac point. From Fig. \ref{figure:diffusive} (g)
we see that, the correctly defined REC drops abruptly
from 1 fs to 5 fs and much more slowly when $t>10$ fs.
While the latter slow decay is a sign of weak localization,
which is usually a precursor of strong localization,
the earlier fast decay cannot be attributed to
a localization effect. Thus, the peak value around 1 fs
(corresponding to a length of about 1 nm) cannot be
taken as the value of $\sigma_{\textmd{sc}}(E)$.
Alternatively, we define $\sigma_{\textmd{sc}}(E)$
as the average value over an appropriate time block
$t_1 \leq t \leq t_2$:
\begin{equation}
\label{equation:sigma_sc_average}
\sigma_{sc}(E) = \frac{1}{t_2-t_1} \int_{t_1}^{t_2}
                 \sigma^{\textmd{GK}}(E, t)dt.
\end{equation}
The time block should be chosen to represent the plateau to which $ \sigma^{\textmd{GK}}(E, t)$ 
saturates to before the onset of localization.
This kind of averaging has been widely used in the
study of thermal conductivity using the  
Green-Kubo method \cite{fan2013,schelling2002}.
Using this alternative definition, the calculated
$\sigma_{\textmd{sc}}(E)$ (the dashed line in 
Fig. \ref{figure:diffusive_sigma})
does not show a peak value
around the Dirac point, and is consistent with that
obtained by Eq. (\ref{equation:sigma_sc_maximum})
in the range of $|E|>0.25$ eV.

The existence of the peak for 
semi-classical conductivity
is also not supported by the work of
Ferreira \textit{et al.} \cite{ferreira2011}.
They directly evaluate the Kubo-Greenwood
formula (Eq. (2)) by expanding both of the $\delta$-functions
using the KPM \cite{ferreira}. Since the KPM is equivalent to the FTM,
as demonstrated earlier,
their method is also equivalent to Fourier transforming
both of the $\delta$-functions,
\begin{equation}
\sigma^{\textmd{KG}}(E) \approx \frac{e^2(\Delta \tau)^2}{2\pi \hbar \Omega}
\sum_{n=-N_m}^{N_m}\sum_{k=-N_m}^{N_m} w_n w_k e^{iE(n+k)\Delta\tau/\hbar}
F^{\textmd{VAC}}_{nk},
\end{equation}
\begin{equation}
F^{\textmd{VAC}}_{nk} \approx
\langle \phi| U(n\Delta \tau) V U(k\Delta\tau) V|\phi\rangle,
\end{equation}
which is in turn equivalent to
applying an extra window function on the VAC
before integrating it up to a given
correlation time (proportional to $N_m$)
in the Green-Kubo formalism.
The extra window function (or damping factor, in the 
context of the KPM) suppresses the localization effect 
and this direct method provides a more unambiguous way
of determining the semi-classical conductivity.
Our new definition of $\sigma_{\textmd{sc}}(E)$ 
is more or less equivalent
to this direct method.

\subsection{The localized transport regime}

Although the Green-Kubo formula
and the Einstein formula have been demonstrated
to be equivalent, we should point out that the Green-Kubo formula
is not practical in the localization regime, for the reason
presented below.
To obtain the REC by integrating the VAC, the time step
should be very small; otherwise, the integration cannot be
accurately evaluated with even very small fluctuations in
the VAC data. However, observing localization requires a
very long total correlation time, and a large number of steps
when using a small time step. At each time step, we need to
calculate the $\delta$ function, which is very time-consuming.
Thus, the necessity of using a small time step in the Green-Kubo
formula makes it impractical in the localized regime.
This probably explains why the results obtained by 
integrating (or summing) the VAC show no evidence of
localization even for a relatively high ($5\%$) 
level of resonant disorder
\cite{yuan2010,ferreira2011}.
In contrast, the Einstein formula is more suitable for studying
the localization behavior, since the numerical evaluation
of the derivative-based REC
does not require a small time step. We thus only use the
Einstein formula in the following discussions of localization.

We begin with a comparison of the results obtained by the Einstein
formula with those by the non-equilibrium
Green's function (NEGF) method \cite{datta1995}.
To our knowledge, a serious
comparison of the two methods in the strongly localized regime
is still absent. We consider AGNRs with a fixed width ($W = 12$ nm)
and a defect concentration of $n=1\%$.
In the NEGF method, the lengths are set by imposing two
conducting leads along the transport direction. In the
Einstein formula, we take a sample size of $95\times 32768$
(which is long enough to eliminate any finite size effect
in the transport direction) and calculate the lengths by
Eq. (\ref{equation:length}).

\begin{figure}
\begin{center}
  \includegraphics[width=3 in]{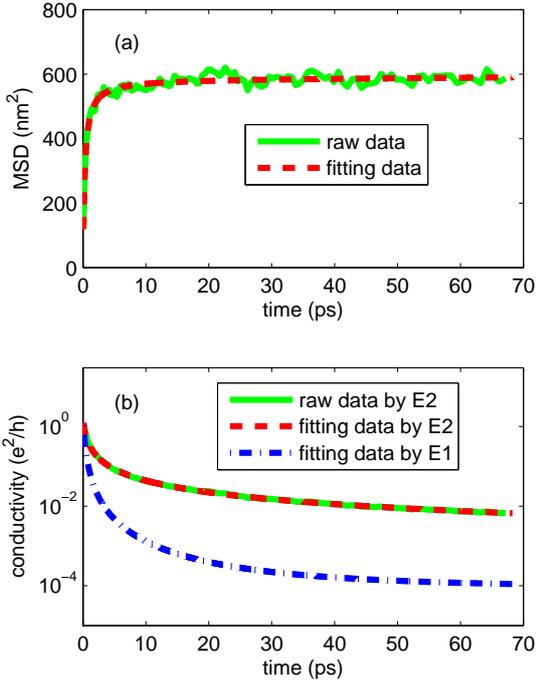}
  \caption{(Color online) Transport properties of  
            AGNR of size $95 \times 32768$ (using 12 random vectors)
            with defect concentration $n=1\%$ at $E = 0.3$ eV.
            (a) Raw and fitted MSD as a function of
            correlation time. (b) Conductivity as a function
            of correlation time.
            Data marked by E1 and E2 correspond to $\sigma^{E1}(E, t)$
            defined by Eq. (9) and $\sigma^{E2}(E, t)$
            defined by Eq. (11), respectively.
            For $\sigma^{E2}(E, t)$, results calculated from the
            raw and the fitted MSD are compared.
            For $\sigma^{E1}(E, t)$, only the results calculated
            from the fitted MSD are presented; those from the raw
            MSD have large fluctuations and cannot be shown completely
            in the same plot.}
  \label{figure:localized_gnr_fit}
\end{center}
\end{figure}

Due to the efficiency of our GPU implementation, we can explore
the strongly localized regime by cheaply calculating the correlation function
up to hundreds of picoseconds for the first time,
eventually observing the saturation of the MSD.
When the MSD saturates, small fluctuations
of the MSD can cause large fluctuations of the REC,
$\sigma^{\textmd{E1}}(E, t)$. Fortunately, we note that
the later part of the MSD can be fitted very well by a
Pad\'{e} approximant of order $[m/n]$:
\begin{equation}
 \Delta X^2(E, t) = \frac{\sum_{j=0}^{m} a_j t^j}
                         {1 + \sum_{k=0}^{n} b_k t^k}.
\end{equation}
Usually, $m = n = 2$ is enough to obtain a good fitting.
An example of the fitting is shown in
Fig. \ref{figure:localized_gnr_fit} (a) for
the energy $E = 0.3$ eV.

Without fitting, the REC $\sigma^{\textmd{E1}}(E, t)$
calculated by Eq. (\ref{equation:REC_E1})
can even develop negative values.
In contrast, the REC $\sigma^{\textmd{E2}}(E, t)$
calculated by Eq. (11) exhibits a very smooth behavior 
even by using the raw data of the MSD
(Fig. \ref{figure:localized_gnr_fit} (b)).
In fact, there is no noticeable difference between the fitted
and the raw data when using the division-based definition
$\sigma^{\textmd{E2}}(E, t)$.
However, in the strongly localized regime where
$\sigma \ll e^2/h$, the two definitions can lead
to a difference of several orders of magnitude for the conductivity
(Fig. \ref{figure:localized_gnr_fit} (b)).

\begin{figure*}
\begin{center}
  \includegraphics[width=3 in]{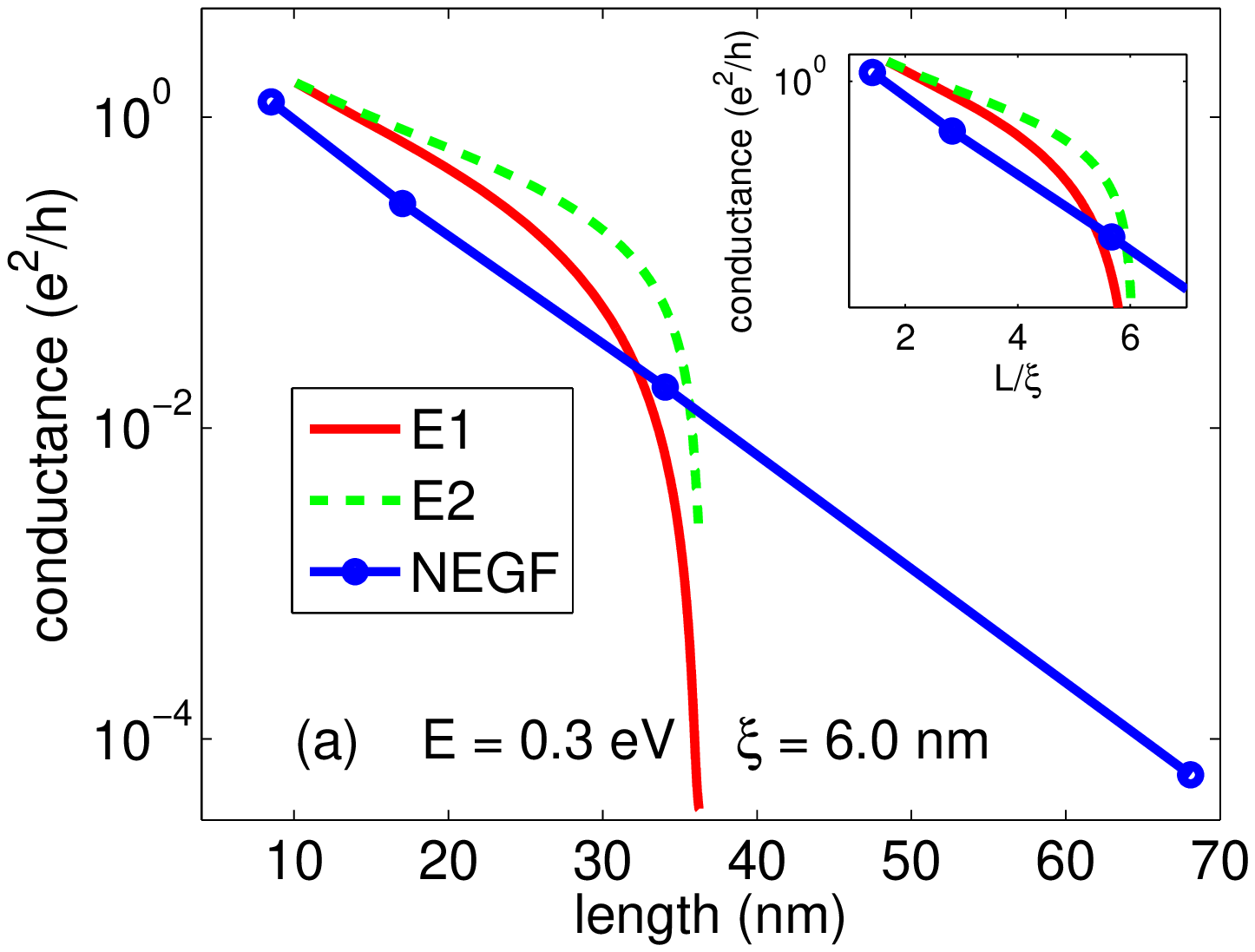}
  \includegraphics[width=3 in]{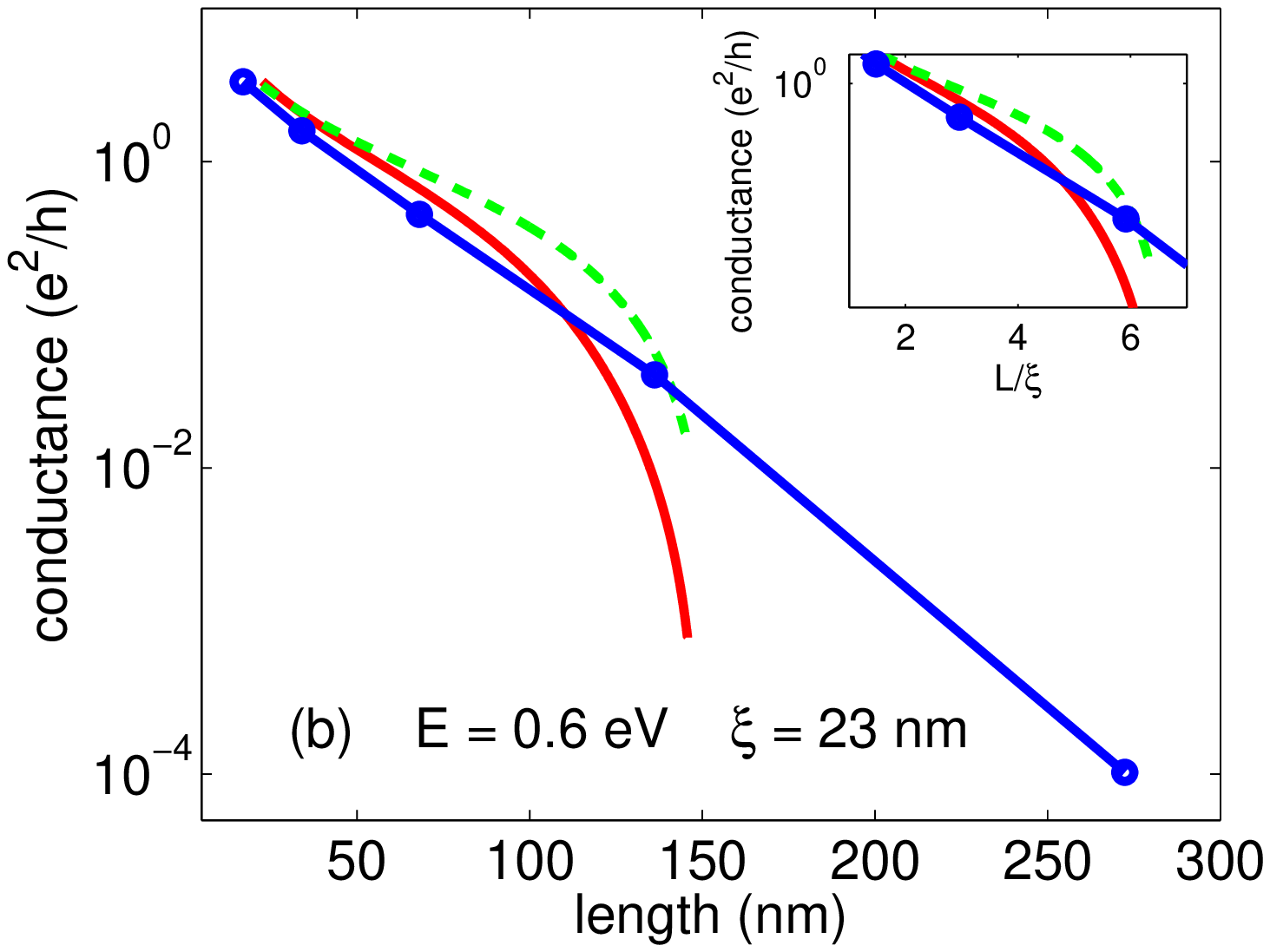}\\
  \includegraphics[width=3 in]{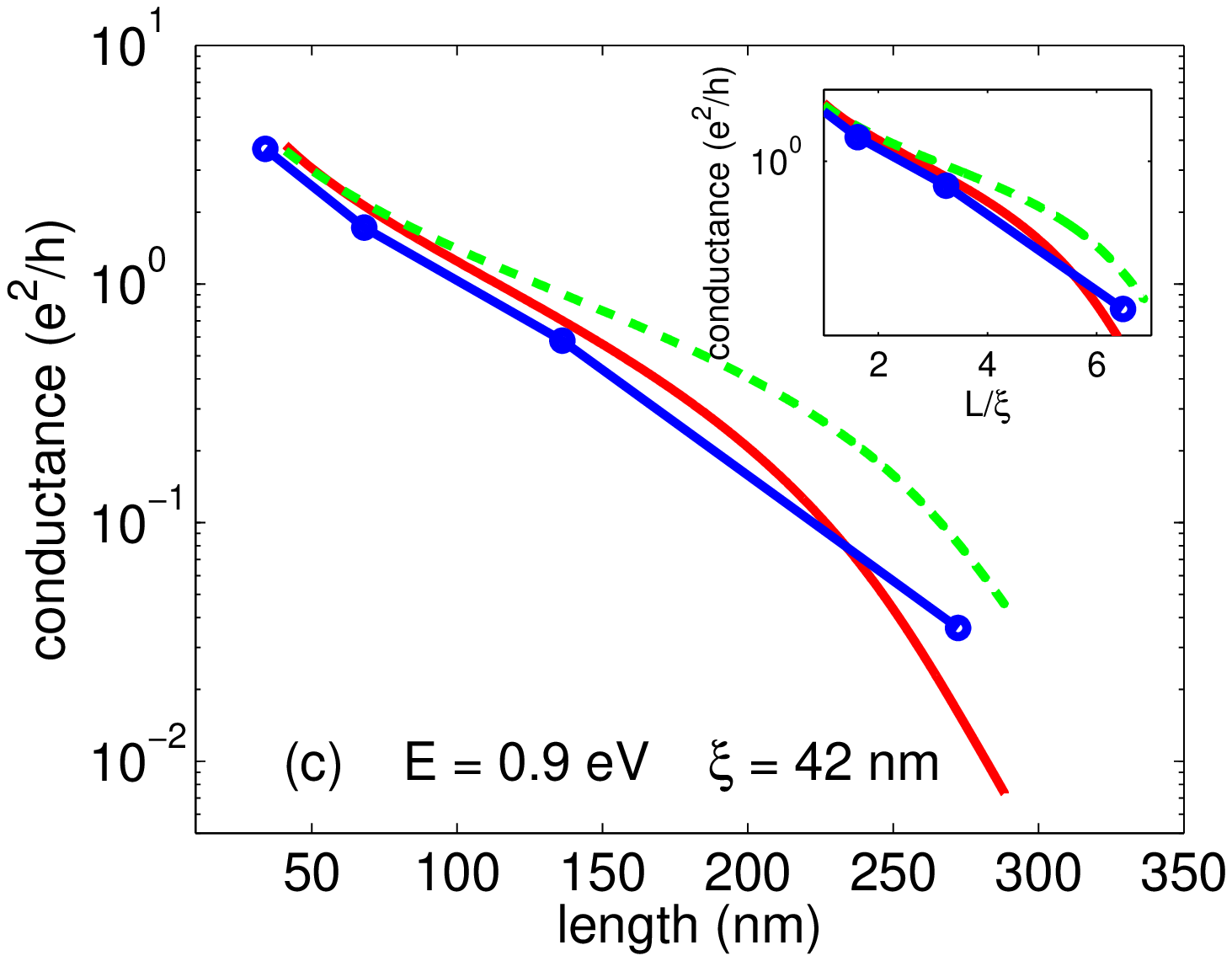}
  \includegraphics[width=3 in]{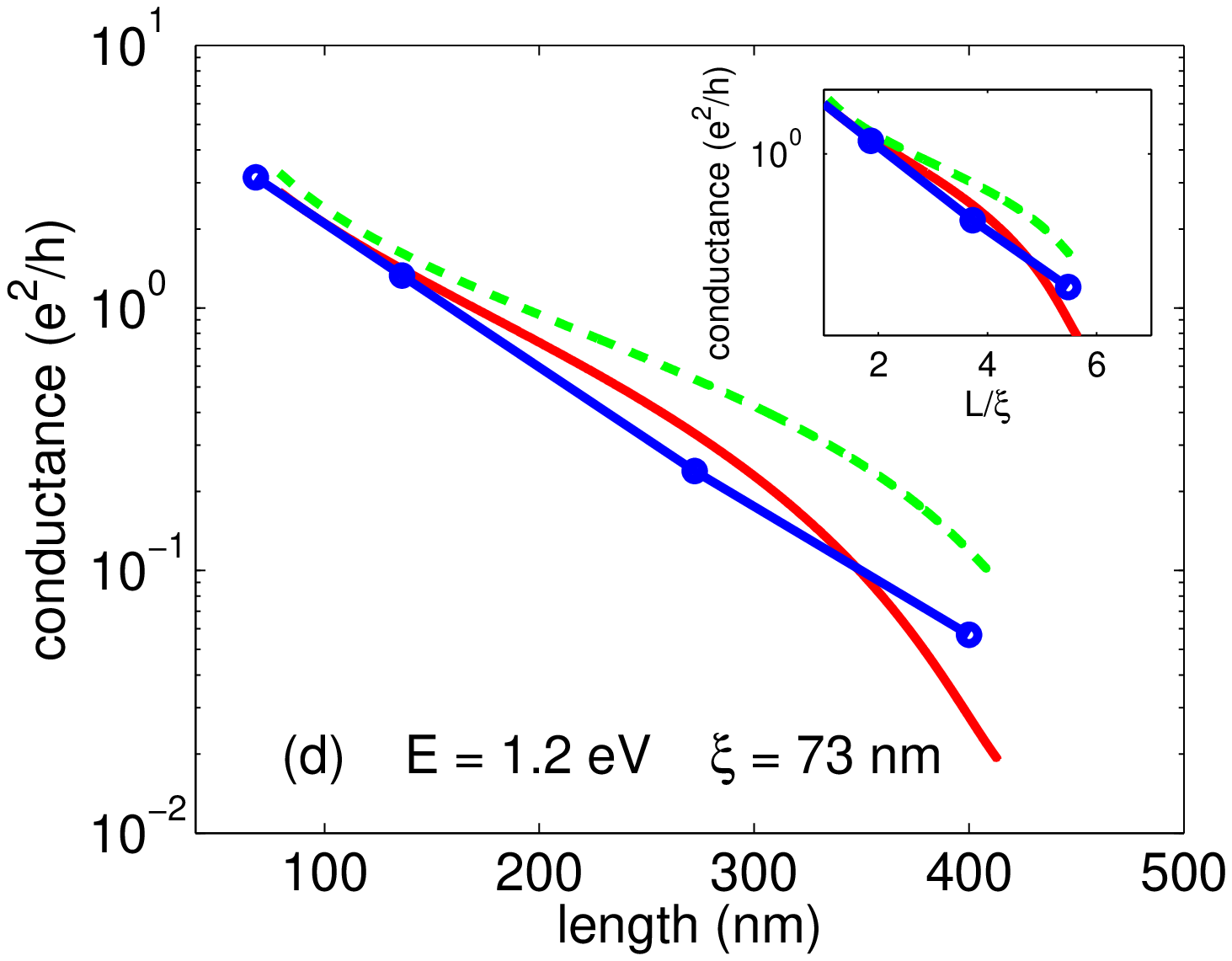}
  \caption{(Color online) Conductance as a function of length
            $L=2 \sqrt{\Delta X^2(E, t)}$ for AGNR of size
            $95 \times 32768$ (using 12 random vectors) 
            with defect concentration $n=1\%$
            at different energies. Solid lines, dashed lines and
            lines with symbols correspond to the results obtained by
            Eq. (9), Eq. (11) and NEGF calculations, respectively.
            The insets show the conductances as a function of
            the reduced length $L/\xi(E)$.}
  \label{figure:localized_gnr_conductance}
\end{center}
\end{figure*}

With a reliable fitting method for obtaining smooth curves
of the MSD and the REC, we can give a quantitative comparison
of the length-dependent conductances as calculated by
\begin{equation}
\label{equation:conductance_localized}
 G(E, L) = \frac{W}{L(E, t)} \sigma(E, t)
\end{equation}
with those calculated by the NEGF method, as shown
in Fig. \ref{figure:localized_gnr_conductance}.
In the NEGF method, the typical conductance \cite{anderson1980}
\begin{equation}
\label{equation:transmission_typical}
 G_{\textmd{typ}}(E, L) = e^{\langle \ln G(E) \rangle}
\end{equation}
is used to represent the ensemble average over $10^2-10^3$
realizations of the defects. As expected,
the conductances calculated by the NEGF method
decay exponentially with the sample length \cite{anderson1980, uppstu2012}
\begin{equation}
\label{equation:conductance_decay}
 G_{\textmd{typ}}(E, L) = G_0(E) e^{-L/\xi(E)},
\end{equation}
where $\xi(E)$ is the localization length and $G_0(E)$ the number of transport modes in the ribbon
multiplied by the conductance quantum $e^2/h$.
The conductances calculated by the Einstein formula
also exhibit an exponential decay up to
$G(E, L) \approx 0.1 e^2/h$.
Within this range,
the correct definition of the REC, $\sigma^{\textmd{E1}}(E, t)$,
results in a very good agreement between the Einstein
formula and the NEGF method.
However, for $G(E, L) < 0.1 e^2/h$, the Einstein formula
fails to capture the length-dependence of the conductance
by using either definition of the REC. In this strongly
localized regime, the conductances calculated by the Einstein
formula decay ``super-exponentially'' with increasing length.

A better characterization of the range within which the Einstein
formalism and the NEGF method give consistent results can be
obtained by plotting the conductances as a function of 
the reduced length $L/\xi(E)$, where the localization length
$\xi(E)$ is deduced from the NEGF results. The length definition
$L(E, t) = 2 \sqrt{\Delta X^2(E, t)}$ in the Einstein formalism
can only be trusted within this range.
As shown in the insets of Fig. \ref{figure:localized_gnr_conductance},
this range can be determined to be $L/\xi(E)<4$, independent of
the energy.

This discrepancy puts the definition of length in the Einstein
formalism into question. Indeed, as seen from
Fig. \ref{figure:localized_gnr_fit}, the MSD will finally
saturate with increasing correlation time, which means
that the length defined in Eq. (\ref{equation:length})
does not increase after the saturation. Thus, the maximum
length that can be probed by the Einstein formula is
bounded from above. In fact, by solving
Eq. (\ref{equation:length}), 
Eq. (\ref{equation:conductance_localized}),
Eq. (\ref{equation:REC_E1}) and
Eq. (\ref{equation:conductance_decay}) simultaneously,
we can get analytical expressions for the length and the MSD:
\begin{equation}
 L(E, t) = 2 \sqrt{\Delta X^2(E, t)}
 = L_0(E) \ln \left(\frac{t + t_1}{t_2} \right),
\end{equation}
where $t_1$ and $t_2$ are two positive parameters
depending on the energy, and $L_0(E)$ is an
energy-dependent length
parameter. However, our simulation results
do not support this solution: the calculated MSD
saturates much faster than logarithmically.
Conceptually, one unambiguous way to define the 
length of a simulated sample is
to connect it with two semi-infinite 
leads along the transport direction,
which affect the effective Hamiltonian 
of the sample by adding the
``self energies'' arising from the interactions 
between the sample and the leads.
This inevitably leads to the ``mesoscopic Kubo-Greenwood formula''
\cite{fisher1981,verges1999,nikolic2001}, or equivalently, the
NEGF method \cite{datta1995}.

\section{Conclusions}
\label{section:Conclusion}

In summary, we have developed an efficient
quantum transport simulation code fully implemented on the GPU,
which attains a speedup factor of 16
(using double-precision) compared
with an optimized serial CPU code.
This seemingly relatively small speedup factor
is obtained by considering the simplest tight-binding
model for graphene, with only three off-diagonal elements
in each row (or column) of the Hamiltonian. We expect that
much higher speedup factors can be obtained when considering
more complicated tight-binding models.
Only electronic transport has been considered in this work; extension
of our GPU implementation to thermal transport \cite{li2011}
should be straightforward and a higher
acceleration rate can be expected due to
the higher computational intensity resulting
from the denser phonon Hamiltonian.
Our methods can also be extended to study other
properties such as local density of states \cite{schubert2009},
which serves an alternative method for studying
Anderson localization. For the interested reader, 
our GPU code is available upon request.

Starting from the Kubo-Greenwood formula,
we have presented a unified picture of the Green-Kubo formula
based on the velocity auto-correlation and the Einstein
formula based on the mean square displacement for DC electrical conductivity
and demonstrated their equivalence for diffusive transport.
We also compared the
kernel polynomial method and the Fourier transform method
for approximating the $\delta$ function and found that
they can be equally used but the former is more efficient.
The demonstration of the equivalence between the Green-Kubo
and the Einstein formula and that between the 
kernel polynomial method and the Fourier transform method
validates our implementation non-trivially.

Using the developed GPU code, we performed a comprehensive evaluation
on the applicability of the method by studying transport properties of
graphene systems in the ballistic, diffusive and localized
regimes. In all the transport regimes, we found that 
the division-based definition of the conductivity
in the Einstein formalism is not equivalent to
the correct derivative-based definition, 
and should be used with caution.

In the ballistic regime, we justified the definition of length
in the Einstein formalism: $L(E,t)=2 \sqrt{\Delta X^2(E, t)}$, where
$\Delta X^2(E, t)$ is the mean square displacement.
We found that the quantized conductance
for graphene nanoribbons can be accurately calculated 
except for the band edges. Around the band edges, 
the conductance is overestimated.
We pointed out that this overestimation arises from the  
difficulty of correctly calculating the density of states and
the velocity, which are both singular around the band edges.

In the diffusive regime, we proposed a new way of finding the   
semi-classical conductivity and compared it with other
approaches. Especially, we established a connection
between our methods and a method which directly 
evaluates the Kubo-Greenwood formula by expanding
both of the $\delta$ functions using the kernel
polynomial method.
Although the Green-Kubo formula is
equivalent to the Einstein formula in the
diffusive regime, the former is not as practical 
as the latter in the localized regime. The reason is
that the former is based on a time-integration and thus
requires a small time step, while the latter is based
on a time-derivative and does not require a small
time step.

In the localized regime, the Einstein formula can
produce results which are consistent with those obtained by
the NEGF method up to some critical
length, $L < 4 \xi(E)$, where $\xi(E)$ is the 
localization length. Although the definition 
of length can only be trusted when $L < 4 \xi(E)$,
in practice, this is enough to observe the weak-to-strong
localization transition. More work is needed to 
clarify the still controversial topics of Anderson
localization in graphene.

\section*{Acknowledgements}
We thank Aires Ferreira, Aur\'{e}lien Lherbier, 
Stephan Roche, and Shengjun Yuan
for helpful discussions.
This research has been supported by the Academy of Finland through
its Centres of Excellence Program (project no. 251748).


\begin{thebibliography}{99}

\bibitem{datta1995}
S. Datta,
Electonic Transport in Mesoscopic Systems.
Cambridge University Press, 1995.

\bibitem{kubo1957}
R. Kubo,
Statistical-Mechanical Theory of Irreversible Processes.
I. General Theory and Simple Applications to Magnetic and Conduction Problems,
J. Phys. Soc. Jpn. \textbf{12}, (1957) 570-586.

\bibitem{greenwood1958}
D. A. Greenwood,
The Boltzmann Equation in the Theory of Electrical Conduction in Metals.
Proc. Phys. Soc. \textbf{71}, (1958) 585-596.

\bibitem{geim2007}
A. K. Geim and K. S. Novoselov,
The rise of graphene,
Nature Materials, \textbf{6}, (2007) 183-191.

\bibitem{neto2009}
A. H. Castro Neto, F. Guinea, N. M. R. Peres, K. S. Novoselov and A. K. Geim,
The electronic properties of graphene,
Rev. Mod. Phys., \textbf{81}, (2009) 109-162.

\bibitem{sancho1985}
M. P. L. Sancho, J. M. L. Sancho and J. Rubio,
Highly convergent schemes for the calculation
of bulk and surface Green functions,
J. Phys. F: Met. Phys, \textbf{15} (1985) 851-858.


\bibitem{mayou1988}
D. Mayou,
Calculation of the Conductivity in the Short-Mean-Free-Path Regime,
Europhys. Lett. \textbf{6}, (1988) 549-554.

\bibitem{mayou1995}
D. Mayou and S. N. Khanna,
A Real-Space Approach to Electronic Transport,
J. Phys. I Paris \textbf{5}, (1995) 1199-1211.

\bibitem{roche1997}
S. Roche and D. Mayou,
Conductivity of quasiperiodic systems: a numerical study,
Phys. Rev. Lett. \textbf{79}, (1997) 2518-2521.

\bibitem{triozon2002}
F. Triozon, J. Vidal, R. Mosseri, and D. Mayou,
Quantum dynamics in two- and three-dimensional quasiperiodic tilings,
Phys. Rev. B \textbf{65}, (2002) 220202(R).

\bibitem{triozon2004}
F. Triozon, S. Roche, A. Rubio, and D. Mayou,
Electrical transport in carbon nanotubes: Role of disorder and helical symmetries,
Phys. Rev. B \textbf{69}, (2004) 121410(R).

\bibitem{markussen2006}
T. Markussen, R. Rurali, M. Brandbyge, and A.-P. Jauho,
Electronic transport through Si nanowires: Role of bulk and surface disorder,
Phys. Rev. B \textbf{74} (2006) 245313.

\bibitem{ishii2010}
H. Ishii, N. Kobayashi, and K. Hirose,
Order-$N$ electron transport calculations from ballistic to diffusive regimes
by a time-dependent wave-packet diffusion method: Application to transport
properties of carbon nanotubes,
Phys. Rev. B \textbf{82}, (2010) 085435.

\bibitem{lherbier2008a}
A. Lherbier, B. Biel, Y.-M. Niquet, and S. Roche,
Transport Length Scales in Disordered Graphene-based Materials: 
Strong Localization Regimes and Dimensionality Effects,
Phys. Rev. Lett. \textbf{100}, (2008) 036803. 

\bibitem{lherbier2008b}
A. Lherbier, X. Blase, Y.-M. Niquet, F. Triozon and S. Roche,
Charge Transport in Chemically Doped Graphene,
Phys. Rev. Lett. \textbf{101}, (2008) 036808.


\bibitem{laissardiere2011}
G. T. de Laissardiere and D. Mayou,
Electronic transport in graphene: quantum effects and role of local defects,
Modern Physics Letters B, \textbf{25} (2011) 1019-1028.



\bibitem{leconte2011}
N. Leconte, A. Lherbier, F. Varchon, P. Ordejon, S. Roche, and J.-C. Charlier,
Quantum transport in chemically modified two-dimensional graphene:
From minimal conductivity to Anderson localization,
Phys. Rev. B \textbf{84} (2011) 235420.


\bibitem{lherbier2012}
A. Lherbier, S. M.-M. Dubois, X. Declerck, Y.-M. Niquet, S. Roche, and J.-C. Charlier,
Transport properties of graphene containing structural defects,
Phys. Rev. B.
\textbf{86}, (2012) 075402.

\bibitem{radchenko2012}
T. M. Radchenko, A. A. Shylau, and I. V. Zozoulenko,
Influence of correlated impurities on conductivity of graphene sheets:
Time-dependent real-space Kubo approach,
Phys. Rev. B \textbf{86}, (2012) 035418.

\bibitem{tuan2013}
D. Van Tuan, J. Kotakoski, T. Louvet, F. Ortmann, J. C. Meyer, and S. Roche,
Scaling properties of charge transport in polycrystalline graphene,
Nano Lett. \textbf{13}, (2013) 1730-1735.


\bibitem{cresti2013}
A. Cresti, F. Ortmann, T. Louvet, D. Van Tuan, and S. Roche,
Broken symmetries, zero-energy modes, and quantum transport in
disordered graphene: from supermetallic to insulating regimes,
Phys. Rev. Lett. \textbf{110}, (2013) 196601.

\bibitem{li2011}
W. Li, H. Sevin\c{c}li, S. Roche, and G. Cuniberti,
Efficient linear scaling method for computing the thermal conductivity
of disodered materials,
Phys. Rev. B \textbf{83}, (2011) 155416.


\bibitem{yuan2010}
S. Yuan, H. De Raedt, and M. I. Katsnelson,
Modeling electronic structure and transport properties of graphene
with resonant scattering centers,
Phys. Rev. B.
\textbf{82}, (2010) 115448.

\bibitem{yuan2010b}
S. Yuan, H. De Raedt, and M. I. Katsnelson,
Electronic transport in disordered bilayer and trilayer graphene,
Phys. Rev. B.
\textbf{82}, (2010) 235409.

\bibitem{yuan2013}
S. Yuan, R. Rold\'{a}n, A.-P. Jauho, and M. I. Katsnelson,
Electronic properties of disordered graphene anditod lattices,
Phys. Rev. B.
\textbf{87}, (2013) 085430.

\bibitem{ari2012}	
A. Harju, T. Siro, F. Federici-Canova, S. Hakala, and T. Rantalaiho,
Computational Physics on Graphics Processing Units,
Lecture Notes in Computer Science, \textbf{7782}, (2013) 3-26.


\bibitem{green1954}
M. S. Green,
Markoff Random Processes and the Statistical Mechanics
of Time-Dependent Phenomena. II. Irreversible Processes in Fluids,
J. Chem. Phys. \textbf{22} (1954) 398-413.


\bibitem{weibe2006}
A. Wei{\ss}e, G. Wellein, A. Alvermann, and H. Fehske,
The kernel polynomial method,
Review of Modern Physics. \textbf{78}, (2006) 275-306.

\bibitem{haydock1972}
R. Haydock, V. Heine and M. J Kelly,
Electronic structure based on the local atomic environment
for tight-binding bands,
J. Phys. C: Solid State Phys. \textbf{5}, (1972) 2845-2858.


\bibitem{haydock1975}
R. Haydock, V. Heine and M. J Kelly,
Electronic structure based on the local atomic environment
for tight-binding bands. II,
J. Phys. C: Solid State Phys. \textbf{8}, (1975) 2591-2605.



\bibitem{feit1982}
M. D. Feit, J. A. Fleck, Jr., and A. Steiger,
Solution of the Schr\"{o}dinger Equation by a Spectral Method,
Journal of Computational Physics, \textbf{47}, (1982) 412-433.


\bibitem{hams2000}
A. Hams and H. De Raedt,
Fast algorithm for finding the eigenvalue distribution of
very large matrices,
Phys. Rev. E. \textbf{62}, (2000) 4365-4377.

\bibitem{ezer1984}
H. Tal-Ezer and R. Kosloff,
An Accurate and Efficient Scheme for Propagating the
Time Dependent Schr\"{o}dinger Equation,
J. Chem. Phys.
\textbf{81}, (1984) 3967-3971.

\bibitem{fehske2009}
H. Fehske, J. Schleede, G. Schubert, G. Wellein, V. S. Filinov, and A. R. Bishop,
Numerical approaches to time evolution of complex quantum systems,
Physics Letters A \textbf{373}, (2009) 2182-2188.

\bibitem{dziubak2012}
T. Dziubak and J. Matulewski,
An object-oriented implementation of a solver of the time-dependent
Schr\"odinger equation using the CUDA technology,
Computer Physics Communications, \textbf{183} (2012) 800-812.

\bibitem{broin2012}
C. \'{O} Broin, and L. A. A. Nikolopoulos,
An OpenCL implementation for the solution of the time-dependent Schrödinger
equation on GPUs and CPUs,
Computer Physics Communications, \textbf{183}, (2012) 2071–2080.





\bibitem{topi2012b}
T. Siro and A. Harju,
Time Propagation of Many-Body Quantum
States on Graphics Processing Units,
Lecture Notes in Computer Science, \textbf{7782}, (2013) 141-152.



\bibitem{cuda}
NVIDIA, CUDA Programming Guide, version 5.0 (2013).


\bibitem{fan2013}
Z. Fan, T. Siro, and A. Harju,
Accelerated molecular dynamics force evaluation on
graphics processing units for thermal conductivity calculations,
Computer Physics Communications, \textbf{184}, (2013) 1414-1425.


\bibitem{topi2012a}
T. Siro, A. Harju,
Exact diagonalization of the Hubbard model on graphics processing units,
Computer Physics Communications, \textbf{183} (2012) 1884-1889.

\bibitem{schelling2002}
P. K. Schelling, S. R. Phillpot, and P. Keblinski, 
Comparison of atomic-level
simulation methods for computing thermal conductivity, 
Phys. Rev. B \textbf{65} (2002) 144306.

\bibitem{ferreira2011}
A. Ferreira, J. Viana-Gomes, J. Nilsson, E. R. Mucciolo, 
N. M. R. Peres, and A. H. Castro Neto,
Unified description of the dc conductivity of monolayer and bilayer
graphene at finite densities based on resonant scatterers,
Phys. Rev. B \textbf{83}, (2011) 165402.

\bibitem{ferreira}
Private communication with A. Ferreira.

\bibitem{anderson1980}
P. W. Anderson, D. J. Thouless, E. Abrahams, and D. S. Fisher
New method for a scaling theory of localization
Phys. Rev. B \textbf{22}, (1980) 3519.


\bibitem{uppstu2012}
A. Uppstu, K. Saloriutta, A. Harju, M. Puska, and A.-P. Jauho,
Electronic transport in graphene-based structures:
An effective cross-section approach
Phys. Rev. B \textbf{85}, (2012) 041401(R).


\bibitem{fisher1981}
D. S. Fisher and P. A. Lee,
Relation between conductivity and transmission matrix,
Phys. Rev. B \textbf{23}, (1981) 6851-6854.


\bibitem{verges1999}
J. A. Verg\'es,
Computational implementation of the Kubo formula for the static conductance:
application to two-dimensional quantum dots,
Computer Physics Communications, \textbf{118}, (1999) 71-80.


\bibitem{nikolic2001}
B. K. Nikoli\'{c},
Deconstructing Kubo formula usage: Exact conductance of a mesoscopic system
from weak to strong disorder,
Phys. Rev. B \textbf{64}, (2001) 165303.


\bibitem{schubert2009}
G. Schubert, J. Schleede, and H. Fehske,
Anderson disorder in graphene nanoribbons: A local distribution approach,
Phys. Rev. B \textbf{79}, (2009) 235116.


\end{thebibliography}
\end{document}